%Paper: hep-th/9503210
%From: Louise Dolan <dolan@augustus.physics.unc.edu>
%Date: Wed, 29 Mar 95 15:42:46 EST

\magnification\magstep1
\tolerance=1600
\parskip = 6pt
\def\dt{{\cdot}}
\def\half{{1\over 2}}
\def\hhalf{\textstyle{1\over 2}}
\def\ref#1{$^{[#1]}$}
\def\pagenumber{\footline={\hss\tenrm\folio\hss}}
\def\nox{{\scriptstyle{\times \atop \times}}}
\nopagenumbers
\rightline{IFP/499/UNC}
\vskip 20pt
\footline={\sevenrm
\hfil$^\ast$ Supported in part by the U.S. Department of
Energy under Grant No. DE-FG 05-85ER40219/Task A\hfil}
\vskip 3pt
\centerline{\bf BRST PROPERTIES OF SPIN FIELDS$^\ast$}
\vskip 50 pt
\centerline {\bf L. Dolan and S. Horvath}
\vskip 12pt
\centerline{\it
Department of Physics and Astronomy,University of North Carolina}
\centerline{\it Chapel Hill, North Carolina 27599-3255, USA}
\vskip 50pt
\centerline {\bf ABSTRACT}
\vskip 6 pt

{\rightskip=18 true mm \leftskip=18 true mm  \noindent
For the closed superstring, spin fields and
bi-spinor states are defined directly in four spacetime dimensions,
in van der Waerden notation (dotted and undotted indices). Explicit operator
product expansions are given, including those for the internal
superconformal field
theory, which are consistent with locality and BRST invariance for the string
vertices. The most general BRST picture changing for these fields is computed.
A covariant notation for the spin decomposition of these
states is developed in which non-vanishing polarizations are selected
automatically. The kinematics of the three-gluon dual model amplitude
in both the Neveu-Schwarz and Ramond sectors in the Lorentz gauges
is calculated and contrasted. Modular invariance and enhanced gauge
symmetry of four-dimensional
models incorporating these states is described.
}
\vfill
\centerline{December 1994}
\vskip12pt
\vfill\eject
\pagenumber
\centerline{\bf1. Introduction}
\vskip5pt
%locality, 4 spacetime dimensions, spin fields, covariant spin decomposition,
%non-abelian coupling, internal cft non-hermitian supercurrent
Conformal spin fields corresponding to physical states in the Ramond sector
of the superstring are analyzed directly in four spacetime dimensions.
BRST properties of spin fields and superconformal fields, associated
with states in the Neveu-Schwarz sector, are compared.
In sect. 2, we describe the conditions on the worldsheet supercurrent
imposed by BRST invariance of the vertex operators and physical states.
In sect. 3, we give a convenient notation for massless spinor states.
This is
useful in discussing a covariant spin decomposition of bi-spinor states and
their relation to circularly polarized
polarization vectors. These are then used to describe
the spin decomposition of the general graviton tensor, and
the vector-spinor states' decomposition into gravitinos and
massless spin-${1\over 2}$ states.
Examples of the bi-spinor states are mesons from the Ramond-Ramond sector
of the Type II superstring.
In sect. 4, operator product expansions  which maintain the locality of
the massless fermion emission vertices are set up for spacetime and
internal spin fields. For these conventional vertices,
BRST picture changing,
including the superconformal ghost number picture $-{3\over 2}$,
is computed; and its dependence on the choice of supercurrent
demonstrated. Tree amplitudes are calculated, and
amplitudes involving Ramond states have the property that
non-vanishing polarizations are selected automatically.
In sect.5, the kinematics of three-point amplitudes of massless particles
are analyzed, leading to the identification of
couplings for Neveu-Schwarz and Ramond mesons.
In sect. 6, we present a set of tree amplitudes involving massless
gauge bosons coming from the Ramond-Ramond sector, which exhibit
a non-abelian structure. We suggest a possible mechanism involving
a non-hermitian piece of the internal worldsheet supercurrent to include
these states in a superstring model satisfying perturbative spacetime
unitarity.
In sect. 7 we discuss a modular invariant partition function.
Conclusions are contained in sect. 8. There are two appendices
concerned with the
two-dimensional sigma matrices used in the Weyl/van der Waerden
description of the spacetime gamma matrices and with the construction
of higher-dimensional gamma matices in the Weyl representation.

\vskip15pt
\centerline{\bf2. Ramond states}
\vskip5pt
We consider the vertex operator directly in four dimensions
$$V_{-{3\over 2}}(k,z) = v^{\dot\alpha}(k) S_{\dot\alpha}(z) e^{ik\dt X(z)}
{\cal S} (z) e^{-{3\over 2}\phi(z)}\,.\eqno(2.1)$$
A suitable choice of supercurrent is given by:
$$F(z) = a_\mu(z) h^\mu(z) + \bar F(z)\,\eqno(2.2)$$
where $0\le\mu\le 3$ and $\bar F(z)$ corresponds to internal degrees of
freedom.
The vertex operator for the Ramond states in the canonical $q=-\half$
superconformal ghost picture is given by picture changing\ref{1,2}
for $k\dt{1\over\sqrt 2}\gamma v \sim u$
where $k\dt{1\over\sqrt 2}\gamma u =0$ by
$$\eqalignno{V_{-\half}(k,\zeta) & = \lim_{z\rightarrow\zeta} e^{\phi(z)} F(z)
V_{-{3\over 2}}(k,\zeta)\cr
&=[u^\alpha (k) S_\alpha (\zeta)
e^{ik\dt X(\zeta)} {\cal S}(\zeta)\cr
&\hskip 8pt + \lim_{z\rightarrow\zeta} (z-\zeta)^{3\over 2} \bar F(z)
v^{\dot\alpha}(k) S_{\dot\alpha} (\zeta) e^{ik\dt X(\zeta)} {\cal S}(\zeta)]
e^{-\half\phi(\zeta)}\,.&(2.3)\cr}$$
BRST invariance of a vertex operator requires its commutator with
the BRST charge $Q$ to be a total divergence; for a vertex operator in
the $q=-{\textstyle{3\over 2}}$ superconformal ghost picture such as (2.1),
this invariance is assured whenever
its operator product with the supercurrent has at most a
$(z-\zeta)^{-{3\over 2}}$ singularity.
We see this as follows:

The ghost superfields\ref{1-3} are
$B(z) = \beta (z) + \theta b(z)$ and
$C(z) = c(z) +\theta\gamma (z)$ with conformal spin $h_\beta = {3\over 2}$,
$h_c = -1$. Then $h_b = 2$ and
$h_\gamma  = -{1\over 2}$. The modings on the Ramond sector and the
commutation relations are
$\{b_n, c_m\} = \delta_{n,-m}$, $[\beta_n, \gamma_m] = \delta_{n,-m}$,
for $n,m\in {\cal Z}$.
Normal ordering is defined by putting the annihilation operators
$b_n$ for $n\ge -1$, $c_n$ for $n\ge 2$ to the right of
the creation operators $b_n$ for $n\le -2$, $c_n$ for $n\le 1$; then
$$b(z)c(\zeta) =
\nox b(z)c(\zeta)\nox
+ {1\over {z-\zeta}}\qquad ;\qquad
c(z)b(\zeta) =
\nox c(z)b(\zeta)\nox + {1\over {z-\zeta}}\,.\eqno(2.4)$$
This is a natural definition for normal ordering as
the ``vacuum'' expectation value of this normal ordered product
including its finite part is zero.
For the superconformal ghosts, normal ordering is defined analogously so that
$$\beta(z)\gamma(\zeta) =
\nox \beta(z)\gamma(\zeta)\nox
- {1\over {z-\zeta}}\qquad ;\qquad
\gamma(z)\beta(\zeta) =
\nox \gamma(z)\beta(\zeta)\nox + {1\over {z-\zeta}}\,.\eqno(2.5)$$
In order to make contact with the field $\phi(z)$ appearing in (2.1),
we can then
``bosonize'' this system, i.e. write it as a theory where operators are
associated with vectors $q$ in the weight lattice of some algebra.
For the bosonic $\beta,\gamma$ conformal field theory,
we define the boson fields
$$\phi(z) \phi(\zeta) = :\phi(z) \phi(\zeta): - {\rm ln}
(z-\zeta)\eqno(2.6)$$
so that
$$:e^{\phi(z)}: :e^{\phi(\zeta)}:\, =\,
:e^{\phi(z)}\,e^{\phi(\zeta)}:\, (z-\zeta)^{-1}\,.\eqno(2.7)$$
Because $:e^{\phi(z)}:$ is a fermion field and $\beta(z)$,$\gamma(z)$ are
bosons, an additional bosonic field $\chi(z)$ is introduced and
$$\eqalignno{\gamma(z) &= :e^{\phi(z)}::e^{-\chi(z)}: \quad;
\qquad J(z) = -\nox \beta\gamma\nox =
\partial\phi\cr
\beta(z) &= :e^{-\phi(z)}::\partial e^{\chi(z)}:&(2.8)\cr}$$
Here
$$\chi(z) \chi(\zeta) = :\chi(z) \chi(\zeta): + \,\eta^{\mu\nu}\,{\rm ln}
(z-\zeta)\,.\eqno(2.9)$$
Because the $\beta\gamma$ spectrum is unbounded from below, it is
useful to define an infinite number of $\beta\gamma$ `vacua'
$|q\rangle_{\beta\gamma}$ where
$\beta_n |q\rangle = 0\,, n > -q - {3\over 2}$,
$\gamma_n |q\rangle = 0\,, n\ge q + {3\over 2}$,
where
$|q\rangle_{\beta\gamma} = e^{q \phi (0)} |0\rangle_{\beta\gamma}$ and
$L_0^{\beta\gamma} |q\rangle = -\half q(q+2) |q\rangle$.
The bosonic $\beta$,$\gamma$ ghost system has two sectors: one is Neveu-Schwarz
where $q\in {\cal Z}$, and the fields $\beta (z)$, $\gamma(z)$, and
$:e^{\phi(z)}:$ are periodic, i.e. half-integrally moded;
the other sector is Ramond, where $q\in {{\cal Z} +\half}$,
and the fields $\beta (z)$, $\gamma(z)$, and
$:e^{\phi(z)}:$ are anti-periodic, i.e. integrally moded.
We note that the conformal fields $:e^{q\phi(z)}:$ for $q$ odd
have the same periodicity and statistics as the supercurrent.
Their conformal dimensions given by
$L_0^{\beta\gamma} |q\rangle = -\half q(q+2) |q\rangle$ are
$\half$ for $q=-1$; $-{3\over 2}$ for $q=1,-3$; and
$-{15\over 2}$ for $q=3,-5$; etc.

The superVirasoro ghost representation has central charge $c=-15$:
$$\eqalignno{
L(z) &= - 2\nox b\partial c\nox - \nox (\partial b) c\nox
- {\textstyle{3\over 2}} \nox\beta{\partial\gamma}\nox
-\hhalf \nox(\partial\beta)\gamma\nox\,&(2.10a)\cr
F(z) &= b\gamma  - 3 \beta\partial c
- 2 (\partial\beta) c\,.&(2.10b)\cr}$$
For the $N=1$ worldsheet supersymmetry system, the BRST charge
$Q \equiv {1\over 2\pi i}
\oint dz Q(z)$ is given from the general form
$$Q(z) \sim c (L^{\rm matter}
+\hhalf L^{\rm ghost}) -\gamma\hhalf (F^{\rm matter} + \hhalf F^{\rm ghost})
\eqno(2.11a)$$
by the BRST current
$$\eqalignno{Q(z) &= Q_0(z) + Q_1(z) +Q_2(z)\cr
Q_0(z) &= c L^{X,\psi} - \nox c b\partial c\nox
+ {\textstyle{3\over 2}}{\partial^2 c}
+ c L^{\beta\gamma}
+ \partial({\textstyle{3\over 4}}\nox\gamma \beta \nox )\cr
Q_1(z) &= -\gamma \,\hhalf F^{X,\psi}\cr
Q_2(z) &= -{\textstyle{1\over 4}}\gamma b\gamma \,.&(2.11b)\cr}$$
Here the matter fields $L^{X,\psi}(z)$ and $F^{X,\psi}(z)$ close the
superconformal algebra with $c=15$; and $L^{\rm ghost}$ and
$F^{\rm ghost}$ denoted in (2.11a) are given in (2.10).
{}From the operator product expansion of the BRST current with itself it
follows that
$$Q^2 =\hhalf\{Q, Q\} = 0\,.\eqno(2.12)$$
This signals the conservation of the BRST charge and allows us to make
different gauge choices.
The commutator which vanishes to insure BRST invariance of the physical states
corresponding to (2.1) is
$$[Q\,,\,V_{-{3\over 2}}(k,z)] =
[Q_0\,,\,V_{-{3\over 2}}(k,z)] + [Q_1\,,\,V_{-{3\over 2}}(k,z)]
+ [Q_2\,,\,V_{-{3\over 2}}(k,z)]\,.\eqno(2.13a)$$
By inspection, we find
$$[Q_0\,,\,V_{-{3\over 2}}(k,z)] = {\textstyle{d\over{dz}}} ( c(z)
V_{-{3\over 2}}(k,z) )\,;\qquad
[Q_2\,,\,V_{-{3\over 2}}(k,z)] = 0\,\eqno(2.13b)$$
and
$$\eqalignno{Q_1(z) V_{-{3\over 2}}(k,\zeta) &= -\hhalf :e^{-\chi(z)}:
e^{\phi(z)} F(z) V_{-{3\over 2}}(k,\zeta)\cr
&= -\hhalf :e^{-\chi(z)}: (z-\zeta)^{3\over 2} e^{-\half\phi(z)}
F(z) v^{\dot\alpha}(k) S_{\dot\alpha}(z) e^{ik\dt X(z)} {\cal S}(z)\cr
&= {\rm regular\, terms}\,&(2.13c)\cr}$$
so that $$[Q_1, V_{3\over 2}(k,z)] = 0\,.\eqno(2.13d)$$

\vskip15pt
\centerline{\bf3. Notation for the product of two spinors}
\vskip5pt
In a Weyl representation, the four-dimensional $\gamma$ matrix
Clifford algebra given by $\{\gamma^\mu,\gamma^\nu\} = 2\eta^{\mu\nu}$ with
$\eta^{\mu\nu} = {\rm diag}(-1,1,1,1)$ can be represented as
$$(\gamma^\mu)^A_{\hskip3pt B} =
\left(\matrix{0&(\bar\sigma^\mu)^\alpha_{\hskip3pt\dot\beta}\cr
(\sigma^\mu)^{\dot\alpha}_{\hskip3pt\beta}&0\cr}\right)\,; \qquad
C^{AB} = \left(\matrix{(i\sigma^2)^{\alpha\beta}&0\cr
0&(i\sigma^2)^{\dot\alpha\dot\beta}\cr}\right)\,
\eqno(3.1)$$
where $\sigma^\mu = (\sigma^0,\sigma^i)$ and
$\bar\sigma^\mu = (-\sigma^0,\sigma^i)$
are given by
$\sigma^0\equiv\left(\matrix{1&0\cr 0&1\cr}\right)$
and the Pauli matrices $\sigma^i$. We define
$\gamma^5 = (i\gamma^0\gamma^1\gamma^2\gamma^3)^A_{\hskip3pt B} =
\left(\matrix{(\sigma^0)^\alpha_{\hskip3pt\beta}&0\cr
0&-(\sigma^0)^{\dot\alpha}_{\hskip3pt\dot\beta}\cr}\right)\,.$
The charge conjugation matrices $C^{AB}$ and $(C^{-1})_{AB}$ are tensors
used to raise and lower indices:
$C^{-1}_{AD} (\gamma^\mu)^D_{\hskip3pt B}\equiv
(\gamma^\mu)_{AB}$ and
$C^{BD} (\gamma^\mu)^A_{\hskip3pt D}\equiv
(\gamma^\mu)^{AB}.$
The transpose relation which defines $C^{AB}$ is
$C^{-1}_{AB}(\gamma^\mu)^B_{\hskip 3pt C} C^{CD} =
-(\gamma^{\mu T})_A^{\hskip3pt D}$ and it
implies the matrices $(\gamma^\mu)^{AB}$ and $(\gamma^\mu)_{AB}$ are
symmetric. For expressions involving the sigma matrices, see
Appendix A.

Since we are in the Weyl representation, we can
use van der Waerden notation\ref{4,5} for spinor indices
$1\le\alpha,\dot\alpha\le 2$.
The two linearly independent solutions to the massless
Dirac equation $k\dt\gamma u^\ell (k) = 0$ are now given by solutions of the
Weyl equations
$$k_\mu\sigma^{\mu\dot\alpha}_{\hskip8pt\beta} u^{1\beta} = 0\;\hskip35pt
k_\mu\bar\sigma^{\mu\alpha}_{\hskip8pt\dot\beta} u^{2\dot\beta} = 0\,
\eqno(3.2a)$$
as $$u^{1\beta}(k) = \left(\matrix{k^0 + k^3\cr
k^1 + ik^2\cr}\right) (k^0+k^3)^{-\half}\,;\qquad
u^{2\dot\beta}(k) =\left(\matrix{-k^1 + ik^2\cr
k^0 + k^3\cr}\right) (k^0+k^3)^{-\half}\,.\eqno(3.2b)$$

We define two additional spinors $v^\ell(k)$ by $k\dt\gamma v^\ell \sim u^\ell$
{\it i.e.}
$${\textstyle{1\over\sqrt 2}}
k_\mu\bar\sigma^{\mu\alpha}_{\hskip8pt\dot\beta} v^{1\dot\beta}
= u^{1\alpha}\;\hskip35pt
{\textstyle{1\over\sqrt 2}}
k_\mu\sigma^{\mu\dot\alpha}_{\hskip8pt\beta} v^{2\beta}
= - u^{2\dot\alpha}\,\eqno(3.3a)$$ as
$$v^{1\dot\beta}(k) = \left(\matrix{k^0 + k^3\cr
k^1 + ik^2\cr}\right) (2(k^0)^2 (k^0+k^3))^{-\half}\,;\quad
v^{2\beta}(k) =\left(\matrix{-k^1 + ik^2\cr
k^0 + k^3\cr}\right) (2(k^0)^2 (k^0+k^3))^{-\half}\,.\eqno(3.3b)$$
Note that formally
$v^{1\dot\beta} = {\textstyle{1\over{\sqrt 2}k^0}} u^{1\beta}$
and $v^{2\beta} = {\textstyle{1\over{\sqrt 2}k^0}} u^{2\dot\beta}$.
In (3.2,3.3) we have $k_\mu k^\mu = 0$.
The spin decomposition of the Weyl bispinors into the two helicity states
of the massless vector is as follows:
$$\eqalignno{u^{1\alpha} u^{1\beta}&=
-\epsilon^+_\lambda k_\kappa (\bar\sigma^\lambda
\sigma^\kappa\sigma^2)^{\alpha\beta}\cr
u^{2\dot\delta} u^{2\dot\gamma}&= \hskip8pt
\epsilon^-_\lambda k_\kappa (\sigma^\lambda
\bar\sigma^\kappa\sigma^2)^{\dot\delta\dot\gamma}\,.&(3.4)\cr}$$
We also find that
$$\eqalignno{u^{1\alpha} v^{1\dot\beta}&= \epsilon^+_\lambda
(\bar\sigma^\lambda
\sigma^2)^{\alpha\dot\beta}\sqrt 2\,;\qquad
u^{2\dot\delta} v^{2\gamma}= \epsilon^-_\lambda (\sigma^\lambda
\sigma^2)^{\dot\delta\gamma}\sqrt 2\cr
v^{1\dot\alpha} u^{1\beta}&= \epsilon^+_\lambda (\sigma^\lambda
\sigma^2)^{\dot\alpha\beta}\sqrt 2\,;\qquad
v^{2\delta} u^{2\dot\gamma} = \epsilon^-_\lambda (\bar\sigma^\lambda
\sigma^2)^{\delta\dot\gamma}\sqrt 2\,&(3.5)\cr}$$
where the expressions in (3.5) are defined only up to a gauge transformation
$\epsilon_\lambda\rightarrow\epsilon_\lambda + k_\lambda$.
In the Lorentz gauges defined by $k\cdot\epsilon^\pm(k) = 0$, we have that
$$i\epsilon^{\mu\nu\lambda\rho} \epsilon_\lambda^{\pm} k_\rho =
\pm(\epsilon^{\mu\pm} k^\nu - \epsilon^{\nu\pm} k^\mu)\,
\eqno(3.6)$$
holds generally for the circularly polarized polarization vectors\ref{6}.
The expressions (3.4,3,5) satisfy (3.2a),(3.3a) with use of (3.6).
We note that (3.4) describes only two polarizations since we can show
that
$$\eqalignno{
\epsilon^-_\lambda k_\kappa (\bar\sigma^\lambda
\sigma^\kappa\sigma^2)^{\alpha\beta}&= 0\,;\qquad
\epsilon^+_\lambda k_\kappa (\sigma^\lambda
\bar\sigma^\kappa\sigma^2)^{\dot\delta\dot\gamma} = 0\,.&(3.7)\cr}$$
The proof of (3.7) is as follows:

Consider
$$\eqalignno{\sigma_\mu
\epsilon^-_\lambda k_\kappa (\bar\sigma^\lambda
\sigma^\kappa\sigma^2)^{\alpha\beta}&=
(\epsilon^-_\mu k_\sigma
-\epsilon^-_\sigma k_\mu + i\epsilon^-_\lambda k_\kappa
\epsilon^{\lambda\kappa}_{\hskip6pt\mu\sigma} )
(\sigma^\sigma\sigma^2)^{\alpha\beta} = 0 &(3.8)\cr}$$
which for $\mu = 0$ is the left equation in (3.7).

The spin decomposition of the Weyl bispinors into the two spin zero states is
$$\eqalignno{u^{1\alpha} u^{2\dot\beta}&= k_\kappa (i\bar\sigma^\kappa
\sigma^2)^{\alpha\dot\beta}\cr
u^{2\dot\delta} u^{1\gamma}&= k_\kappa (i\sigma^\kappa
\sigma^2)^{\dot\delta\gamma}\,.&(3.9)\cr}$$

We also note here the spin decomposition of the spin-vector state
$\psi_\mu^A =\epsilon_\mu^+ u^A$ separated
into its spin-${\textstyle{3\over 2}}$
and spin-$\hhalf$ content. This is simple in van der Waerden notation and
eliminates the need for introducing\ref{7,8} a noncovariant momentum
vector $\bar k$ where $k\cdot\bar k = 1$.
We find the spin-${\textstyle{3\over 2}}$ part to be given by
$$\eqalignno{\psi_\mu^{+\alpha} = \epsilon_\mu^+ u^{1\alpha}&\qquad
{\rm helicity}={\textstyle{3\over 2}}\,;\qquad
\psi_\mu^{-\dot\beta} = \epsilon_\mu^- u^{2\dot\beta}\qquad
{\rm helicity}=-{\textstyle{3\over 2}}&(3.10)\cr}$$
since the spin-vectors in (3.10) satisfy
the on-shell $k^2 = 0$ Rarita-Schwinger equation
$$k\cdot\psi^\pm = 0\,;\qquad k\cdot\gamma\psi^\pm_\mu =0\,;\qquad
\gamma\cdot\psi = 0\,.\eqno(3.11)$$
For example, for the spin-${\textstyle{3\over 2}}$ helicity,
$$\gamma\cdot\psi^\pm = \sigma^{\mu\dot\alpha}_{\hskip8pt\alpha}
\epsilon_\mu^+ u^{1\alpha} = 0\,.\eqno(3.12)$$
To prove (3.12), consider
$$\eqalignno{\sigma^{\mu\dot\alpha}_{\hskip8pt\alpha}
\epsilon_\mu^+ u^{1\alpha}v^{1\dot\beta}&=\epsilon_\mu^+
\sigma^{\mu\dot\alpha}_{\hskip8pt\alpha}\epsilon^+_\lambda
(\bar\sigma^\lambda\sigma^2)^{\dot\alpha\dot\beta}{\sqrt 2}\cr
&={\sqrt 2} \epsilon^+\cdot\epsilon^+\sigma^2 = 0\,.&(3.13)\cr}$$
The spin-$\hhalf$ part is
$$\eqalignno{\chi_\mu^{+\dot\beta} = \epsilon_\mu^+ u^{2\dot\beta}&\qquad
{\rm helicity}={\textstyle{1\over 2}}\,;\qquad
\chi_\mu^{-\alpha} = \epsilon_\mu^- u^{1\alpha}\qquad
{\rm helicity}=-{\textstyle{1\over 2}}\,&(3.14)\cr}$$
since for the spin vectors in (3.14) we have
$$k\cdot\chi = 0\,;\qquad k\cdot\gamma\chi_\mu =0\,;\qquad
\gamma\cdot\chi \ne 0\,.\eqno(3.15)$$

The connection between the Lorentz
covariant notation developed here and the $k,\bar k$ notation
used in [7,8] is seen explicitly in (3.16a,b).
As an example, consider the helicity equal to $+\half$ part,
$\chi_\mu^{+\dot\beta} = \epsilon_\mu^+ u^{2\dot\beta}$.
Then
$$\eqalignno{\chi_\mu^{+\dot\beta}&= \epsilon_\mu^+ u^{2\dot\beta}&(3.16a)\cr
&= -{\textstyle{i\over 2}}
(\sigma_\mu - k_\mu\bar k\dt\sigma)^{\dot\beta}_{\hskip3pt\beta} u^{1\beta}
&(3.16b)\cr}$$
so that
$$\eqalignno{\gamma\dt\chi^+ =
\epsilon_\mu^+ \bar\sigma^{\mu\alpha}_{\hskip8pt\dot\beta}
u^{2\dot\beta}&= -i u^{1\alpha}\cr
&= -{\textstyle{i\over 2}} (\bar\sigma^\mu
(\sigma_\mu - k_\mu\bar k\dt\sigma))^{\dot\beta}_{\hskip3pt\beta} u^{1\beta}\,.
&(3.17)\cr}$$
The general graviton tensor $\epsilon_{\mu\nu}$ in this notation is given by:
$$\eqalignno{h_{\mu\nu} &= \hhalf (\epsilon_{\mu\nu} + \epsilon_{\nu\mu})
- {1\over d-2} \epsilon^\rho_{\hskip3pt\rho} (\eta_{\mu\nu} - k_\mu\bar k_\nu
-\bar k_\mu k_\nu ) = \epsilon^+_\mu\epsilon^+_\nu \,{\rm or}\,
\epsilon^-_\mu\epsilon^-_{\nu}\cr
B_{\mu\nu} &= \hhalf (\epsilon_{\mu\nu} - \epsilon_{\nu\mu}) =
\hhalf (\epsilon^+_\mu\epsilon^-_\nu - \epsilon^-_\mu\epsilon^+_\nu )\cr
D_{\mu\nu} &= {1\over {\sqrt {d-2}}} (\eta_{\mu\nu} - k_\mu\bar k_\nu
-\bar k_\mu k_\nu ) = {2\over {\sqrt {d-2}}} (\epsilon^+_\mu\epsilon^-_\nu +
\epsilon^-_\mu\epsilon^+_{\nu})\,.&(3.18)\cr}$$
Here $\epsilon_{\mu\nu} (k)$ is polarization tensor which satisfies
$k^\mu\epsilon_{\mu\nu} (k) = k^\nu\epsilon_{\mu\nu} (k) = 0$,
and represents either a (symmetric traceless) graviton $h_{\mu\nu}$,
an anti-symmetric tensor $B_{\mu\nu}$ which is a pseudo-scalar in
four dimensions, or a (transverse diagonal)
scalar dilaton $D$. We remark that the
right-hand terms in (3.18) provide a covariant spin decomposition
of the spin\nobreak-\nobreak2 tensor
since they are independent of $\bar k$. They
follow from the Lorentz gauge condition $k^\mu\epsilon^\pm_\mu = 0$ and
the normalization conditions
$\epsilon_\mu^+\epsilon^{\mu +} = \epsilon_\mu^-\epsilon^{\mu -} = 0$,
$\epsilon_\mu^+\epsilon^{\mu -} = 1$.
\vskip15pt
\vfill\eject
\centerline{\bf 4. BRST invariance and picture changing}
\vskip 5pt
In four spacetime dimensions, the vertex operators associated with
massless Ramond states can be described by
$$\eqalignno{V^{(1)}_{-{3\over 2}}(k,z) &= v^{1{\dot\alpha}}(k)
S_{\dot\alpha}(z) e^{ik\dt X(z)} f^\ell
\Sigma_\ell (z)\,e^{-{3\over 2}\phi(z)}\cr
V^{(2)}_{-{3\over 2}}(k,z) &= v^{2{\alpha}}(k)
S_{\alpha}(z) e^{ik\dt X(z)}
f^{\dot\ell}\Sigma_{\dot\ell}(z)\,e^{-{3\over 2}\phi(z)}\,&(4.1)\cr}$$
where $v^{1{\dot\alpha}}$, $v^{2{\alpha}}$ are given in (3.3b).
In order to maintain locality for the string vertices, we choose the following
operator product expansions:
$$\eqalignno{S_\alpha(z) S_\beta(\zeta) &= (z-\zeta)^{-\half}
C^{-1}_{\alpha\beta}
%+(z-\zeta)^{\half}\,[\gamma^\mu ,\gamma^\nu]_{\alpha\beta}
%{\textstyle{1\over 4}}
%\psi_\mu(\zeta)\psi_\nu(\zeta)
+\ldots\cr
S_\alpha (z) S_{\dot\beta}(\zeta) &= (z-\zeta)^0 \gamma^\mu_{\alpha\dot\beta}
{\textstyle{1\over\sqrt 2}}\psi_\mu(\zeta) +
\ldots\cr
S_{\dot\alpha} (z) S_{\beta}(\zeta)
&= (z-\zeta)^0 \gamma^\mu_{\dot\alpha\beta}
{\textstyle{1\over\sqrt 2}}\psi_\mu(\zeta) +\ldots\cr
S_{\dot\alpha} (z) S_{\dot\beta}(\zeta)
&= (z-\zeta)^{-\half} C^{-1}_{\dot\alpha\dot\beta}
+\ldots&(4.2a)\cr}$$
$$\eqalignno{\Sigma_\ell(z)\Sigma_{\dot n} (\zeta) &=
(z-\zeta)^{-{3\over 4}} C^{-1}_{\ell\dot n}
+\dots\cr
\Sigma_{\dot\ell} (z)\Sigma_n (\zeta)
&= (z-\zeta)^{-{3\over 4}} C^{-1}_{\dot\ell n}
+\dots\cr
\Sigma_\ell(z)\Sigma_n (\zeta) &=
(z-\zeta)^{-{1\over 4}}\Gamma^a_{\ell n}
{\textstyle{1\over\sqrt 2}}\psi_a (\zeta) +
\ldots\cr
\Sigma_{\dot\ell} (z)\Sigma_{\dot n} (\zeta) &=
(z-\zeta)^{-{1\over 4}}\Gamma^a_{\dot\ell \dot n}
{\textstyle{1\over\sqrt 2}}\psi_a (\zeta) +
\ldots&(4.2b)\cr}$$
where the subleading terms are less singular by integer powers of
$(z-\zeta)$, and
the field $\psi^a(z)$ is understood to represent
$\tilde\gamma^5 \otimes \psi^a(z)$,
in order to maintain anticommutativity of $\psi^a(z)$ and $\psi^\mu (z)$.
The operator relations above are not single-valued, but when taken in
appropriate combinations
with each other and the ghost fields, the resulting string vertices
are local (in the
sense of meromorphic operator product expansions), at
least at zero momentum; and the momentum-dependent fields are local
when the complete
closed string holomorphic and antiholomorphic expressions are considered.

In (4.2b), we have chosen a free fermion form for part of the internal
conformal field theory with $c=3$. Here
$1\le a\le 6$ and we use a Weyl representation for the internal
gamma matrices
given for $1\le a\le 3\,,\,1\le \ell,\dot\ell\le 4$ by
$$\Gamma^a =\left (\matrix{0&(\alpha^a)^\ell_{\hskip3pt\dot m}\cr
-(\alpha^a)^{\dot\ell}_{\hskip3pt m}&0\cr}\right )\,;\,
\Gamma^{a+3} = i \left (\matrix{0&(\beta^a)^\ell_{\hskip3pt\dot m}\cr
(\beta^a)^{\dot\ell}_{\hskip3pt m}&0\cr}\right )\,;
\,C = \left (\matrix{0&(I_4)^{\ell\dot m}\cr
(I_4)^{\dot\ell m}&0\cr}\right ).\eqno(4.3)$$
Also
$\tilde\gamma^5\equiv\gamma^5\,(-1)^{\sum_{n>0}{{\psi^\mu_{-n}\psi^\mu_n}}}$
and
$\tilde\Gamma^7\equiv(i\Gamma^1\Gamma^2\Gamma^3\Gamma^4\Gamma^5\Gamma^6)
(-1)^{\sum_{n>0}\psi^a_{-n}\psi^a_n}$ where
$i\Gamma^1\Gamma^2\Gamma^3\Gamma^4\Gamma^5\Gamma^6 =
\left (\matrix{(I_4)^{\ell}_{\hskip3pt m}&0\cr
0&-(I_4)^{\dot\ell}_{\hskip3pt\dot m}\cr}\right )\,.$
The matrices $\alpha^a$, $\beta^a$ are real antisymmetric.
See Appendix B.
In the Weyl representation,
the charge conjugation and gamma matrices have the following
properties:
\leftline{for four dimensions,}
$$C_{\alpha\beta} = - C_{\beta\alpha}\,,\qquad
\gamma^\mu_{\alpha\dot\beta} = \gamma^\mu_{\dot\beta\alpha}\,,\quad {\rm etc.}
\eqno(4.4a)$$
\noindent for six dimensions,
$$C_{\ell\dot n} =  C_{\dot n\ell}\,,\qquad
\Gamma^a_{\ell n} = - \Gamma^a_{n\ell}\,,\quad {\rm etc.}  \eqno(4.4b)$$
\noindent for ten dimensions,
$$C_{A\dot B}=  - C_{\dot B A}\,,\qquad
\Gamma^M_{AB} = \Gamma^M_{BA}\,,\quad {\rm etc.}  \eqno(4.4c)$$
For a general internal field theory, the expression
$\Gamma^a_{\ell n}
{\textstyle{1\over\sqrt 2}}\psi_a (z)$ in (4.2b) is replaced by
$\psi_{\ell n}(z)$, which has conformal weight $\hhalf$ and is
antisymmetric in $\ell, n$.
To check the locality of the vertex operators in (4.1) at zero
momentum, we use
(4.2,4) to show
$$\eqalignno{V^{(1)}_{-{3\over 2}} (z)\, V^{(1)}_{-{3\over 2}} (\zeta)
&= (z-\zeta)^{-3} v^{1\dot\alpha} (k_1) v^{1\dot\beta} (k_2)
C^{-1}_{\dot\alpha\dot\beta} f^\ell f^n \Gamma_{\ell n}^a
{\scriptstyle{1\over\sqrt 2}}\psi_a(\zeta) e^{-3\phi (\zeta)} +\ldots\cr
&\cong - V^{(1)}_{-{3\over 2}} (\zeta)\,V^{(1)}_{-{3\over 2}}(z)&(4.5a)\cr}$$
where in (4.5a) the operator product
$V^{(1)}_{-{3\over 2}} (z)\,V^{(1)}_{-{3\over 2}} (\zeta)$ is defined
for $|z|>|\zeta|$, the product
$V^{(1)}_{-{3\over 2}} (\zeta)\,V^{(1)}_{-{3\over 2}}(z)$ is defined
for $|\zeta|>|z|$ and $\cong$ denotes equal in the sense of
analytic continuation\ref{1,9}.
Similary, we find local fermionic fields
$$\eqalignno{V^{(i)}_{-{3\over 2}} (z)\, V^{(j)}_{-{3\over 2}} (\zeta)
&\cong - V^{(j)}_{-{3\over 2}} (\zeta)\,V^{(i)}_{-{3\over 2}}
(\zeta)&(4.5b)\cr}$$
for all $1\le i,j\le 2$, using (4.1,4.2).

{}From (2.3), we see that by
selecting a particular supercurrent, we can define $q=-\half$
ghost picture fermion vertex operators.
{}From (2.13c), we know that any supercurrent whose operator product
with a fermion vertex operator has at most a $(z-\zeta)^{-{3\over 2}}$
singularity will ensure BRST invariance for the vertex operator.
In this section, we choose a supercurrent (2.2) such that the internal
$\bar F(z)$ has at most a $(z-\zeta)^{-{1\over 2}}$ singularity with
(4.1). Then from (3.3a),
$$\eqalignno{V^{(1)}_{-\half}(k,\zeta)
&=\lim_{z\rightarrow\zeta} e^{\phi(z)} F(z)
V^{(1)}_{-{3\over 2}}(k,\zeta)\cr
&= u^{1\alpha} (k) S_\alpha (\zeta) e^{ik\dt X(\zeta)} f^\ell \Sigma_\ell
(\zeta)
e^{-\half\phi(\zeta)}\,.&(4.6a)\cr}$$
$$\eqalignno{V^{(2)}_{-\half}(k,\zeta)
& = \lim_{z\rightarrow\zeta} e^{\phi(z)} F(z)
V^{(2)}_{-{3\over 2}}(k,\zeta)\cr
&= -u^{2\dot\alpha} (k) S_{\dot\alpha}
(\zeta) e^{ik\dt X(\zeta)}
f^{\dot\ell}\Sigma_{\dot\ell}(\zeta)
e^{-\half\phi(\zeta)}\,.&(4.6b)\cr}$$
BRST invariance of the vertex operators in the $q=-\half$ picture (4.6) is
assured from BRST invariance of the picture changed $q=-{\textstyle{3\over 2}}$
operators and from (2.12), or can be checked directly.
Locality holds from (4.2) for these operators as well:
$$\eqalignno{V^{(1)}_{-{1\over 2}} (z)\, V^{(1)}_{-{1\over 2}} (\zeta)
&= (z-\zeta)^{-1} u^{1\alpha} (k_1) u^{1\beta} (k_2)
C^{-1}_{\alpha\beta} f^\ell f^n \Gamma_{\ell n}^a
{\scriptstyle{1\over\sqrt 2}}\psi_a(\zeta) e^{-\phi (\zeta)} +\ldots\cr
&\cong - V^{(1)}_{-{1\over 2}} (\zeta)\,V^{(1)}_{-{1\over 2}}(z)&(4.7a)\cr}$$
where we note that unlike the Neveu-Schwarz case, the statisitics
of the fields associated with Ramond states
does not change from one picture to another.
In general, $V^{(i)}_{-{1\over 2}} (z)\, V^{(j)}_{-{1\over 2}} (\zeta)
\cong - V^{(j)}_{-{1\over 2}} (\zeta)\,V^{(i)}_{-{1\over 2}}
(\zeta)$.
Also, locality holds between fields in different pictures:
$$\eqalignno{V^{(1)}_{-{1\over 2}} (z)\, V^{(1)}_{-{3\over 2}} (\zeta)
&= (z-\zeta)^{-1} u^{1\alpha} (k_1) v^{1\dot\beta} (k_2)
\gamma^\mu_{\alpha\dot\beta} \psi_\mu (\zeta)f^\ell f^n \Gamma_{\ell n}^a
\psi_a(\zeta){\scriptstyle{1\over2}} e^{-\phi (\zeta)} +\ldots\cr
&\cong - V^{(1)}_{-{1\over 2}} (\zeta)\,V^{(1)}_{-{1\over 2}}(z)&(4.7b)\cr}$$
To establish (4.7b), we note as previously mentioned below (4.2) that
the field $\psi^a(z)$ is understood to represent
$\tilde\gamma^5 \otimes \psi^a(z)$, and we use $\gamma^5 u^1\sim u^1$,
$\gamma^5 v^1 \sim - v^1$, etc.
In general, $V^{(i)}_{-{1\over 2}} (z)\, V^{(j)}_{-{3\over 2}} (\zeta)
\cong - V^{(j)}_{-{3\over 2}} (\zeta)\,V^{(i)}_{-{1\over 2}}
(\zeta)$.

The vertex operators for the massless Neveu-Schwarz states in the canonical
$q=-1$ superconformal ghost picture are
$$\eqalignno{
V_{-1}(k,z,\epsilon)
&=\epsilon\cdot\psi(z) e^{ik\dt X(z)}\,e^{-\phi(z)}&(4.8a)\cr
V_{-1}^a(k,z) &= \tilde\gamma^5\otimes
\psi^a(z) e^{ik\dt X(z)}\,e^{-\phi(z)}&(4.8b)\,.\cr}$$
States in the Neveu-Schwarz matter system (i.e. without ghosts) form\ref{1}
superconformal fields
$V(z,\theta) = V_q (z) + \theta V_{q+1}(z)$
with upper and lower components related by
$$\eqalignno{G(z) V_q(\zeta) &= (z-\zeta)^{-1} V_{q+1}(\zeta)\cr
G(z) V_{q+1} (\zeta) &= (z-\zeta)^{-2} 2h_q V_q(\zeta) +
(z-\zeta)^{-1}\partial V_q(\zeta)\,.&(4.9)\cr}$$

BRST invariance holds for both the vertices (4.8) since
$$[Q_0\,,\,V_{-1}(k,z)] = {\textstyle{d\over{dz}}}[ c(z)
V_{-1}(k,z) ]\,;\qquad
[Q_2\,,\,V_{-1}(k,z)] = 0\,\eqno(4.10)$$
and
$$\eqalignno{Q_1(z) V_{-1}(k,\zeta) &= -\hhalf :e^{-\chi(z)}:
e^{\phi(z)} G(z) V_{-1}(k,\zeta)\cr
&= -\hhalf :e^{-\chi(z)}: (z-\zeta)^1
G(z) V_{-1}^{\rm matter}(k,\zeta)\cr
&=-\hhalf :e^{-\chi(z)}: V_0^{\rm matter}(k,\zeta)\cr
&= {\rm regular\, terms}\,&(4.11a)\cr}$$
so that $$[Q_1, V_{-1}k,z)] = 0\,.\eqno(4.11b)$$
In the $q=0$ superconformal ghost picture, (4.8) is
$$\eqalignno{
V_{0}(k,z,\epsilon)
&=\lim_{z\rightarrow\zeta} e^{\phi(z)} G(z) V_{-1}(k,\zeta\epsilon)\cr
&= [k\cdot\psi(\zeta) \epsilon\cdot\psi(\zeta) + \epsilon\cdot a(\zeta)]
e^{ik\dt X(\zeta)}\,&(4.12a)\cr
V_{0}^a(k,z) &= \lim_{z\rightarrow\zeta} e^{\phi(z)} G(z)
V_{-1}^a(k,\zeta)\cr
&=[k\cdot\psi(\zeta)
\tilde\gamma^5\otimes\psi^a(\zeta)
+ (\lim_{z\rightarrow\zeta} (z-\zeta) \bar F(z) \tilde\gamma^5\otimes
\psi^a(\zeta)\,)\,]\,
e^{ik\dt X(\zeta)}\,.&(4.12b)\cr}$$
The tree correlation functions of BRST invariant vertex operators are
independent of the distribution of ghost charges given that
$\sum_i q_i = -2$.
As an example we consider the coupling of the {\it general graviton tensor
with two massless
vector mesons}. For Neveu-Schwarz mesons we have
$$\eqalignno{
&\langle 0|V^a_{-1}(k_1,z_1)\,V_0(k_2,z_2,\epsilon_2)\,
V^b_{-1}(k_3,z_3) c(z_1)c(z_2)c(z_3)|0\rangle \cr
&\dt\langle 0|V_{-1}(k_1,\bar z_1,\epsilon_1)\,V_0(k_2,\bar z_2,\epsilon_2)\,
V_{-1}(k_3,\bar z_3,\epsilon_3) c(\bar z_1)c(\bar z_2)c(\bar z_3)|0\rangle \cr
&= \delta^{ab} \epsilon_2^{\mu\nu} k_{3\mu} [k_{3\nu} \epsilon_1\dt\epsilon_3
+ \epsilon_{3\nu} \epsilon_1\dt k_2 + \epsilon_{1\nu} \epsilon_3\dt k_1]
&(4.13)\cr}$$
where conformal ghost contributions\ref{1} are included, such as
$$\langle 0|c(z_1)c(z_2)c(z_3)|0\rangle = (z_1 - z_2) (z_2 - z_3)
(z_1 - z_3)\,.\eqno(4.14)$$
For Ramond-Ramond mesons we have
$$\eqalignno{
&\langle 0|V^{(2)}_{-{1\over 2}}(k_1,z_1)\,V_{-1}(k_2,z_2,\epsilon_2)\,
V^{(1)}_{-{1\over 2}}(k_3,z_3) c(z_1)c(z_2)c(z_3)|0\rangle \cr
&\dt\langle 0|V^{(2)}_{-{1\over 2}}(k_1,\bar z_1)\,V_{-1}(k_2,\bar
z_2,\epsilon_2)\,
V^{(1)}_{-{1\over 2}}(k_3,\bar z_3) c(\bar z_1)c(\bar z_2)c(\bar z_3)
|0\rangle \cr
&= u^{2\dot\delta} (k_1) u^{2\dot\gamma} (k_1)\epsilon_{2\mu}
(-i\sigma^2\sigma^\mu)_{\dot\gamma\alpha} u^{1\alpha} (k_3) u^{1\beta} (k_3)
\epsilon_{2\nu}\,(-i\sigma^2\bar\sigma^\nu)_{\beta\dot\delta}
\hhalf f^{\dot k} f^{\dot n} C^{-1}_{\dot n\ell}
f^\ell f^m C^{-1}_{m\dot k}\cr
&= {\scriptstyle {1\over 8}} \delta^{IJ}
\epsilon_{1\rho}^- k_{1\sigma} \epsilon_{2\mu\nu}
\epsilon^+_{3\lambda} k_{3\kappa}\, {\rm tr}
\,(\sigma^\rho\bar\sigma^\sigma\sigma^\mu\bar\sigma^\lambda\sigma^\kappa
\bar\sigma^\nu\,)\cr
&= -\delta^{IJ}\epsilon_2^{\mu\nu} k_{3\mu}
[k_{3\nu}\epsilon^-_1\dt\epsilon_3^+
+ \epsilon^+_{3\nu} \epsilon_1^-\dt k_2
+ \epsilon^-_{1\nu} \epsilon^+_3\dt k_1]
&(4.15)\cr}$$
which up to a sign is the same as (4.13) for these polarizations.
To evaluate (4.15), we have used a trace formula in Appendix A and
various on-shell identities from sect. 5. In (4.15)
only $\epsilon_{\mu\nu} = h_{\mu\nu}$ survives.
%It is clear from (4.15) how the
%polarization vectors select non-vanishing amplitudes, since
%we see from the index structure on $\gamma^\mu$
%that the tree amplitude for the general graviton tensor
%coupled to Ramond vector mesons with, say, the polarizations
%$\epsilon_1^+,\epsilon_3^+$ vanishes identically.
We normalize the vertex operators by expanding
$f_L^\ell f_R^m = {\textstyle{1\over 4}} M^J_{\ell m}$ where $1\le J\le 16$
and $M_{\ell m}^J$ are a complete set of sixteen linearly independent real
four by four matrices $M^J = \{\alpha^a,\beta^a,\alpha^a\beta^b, I_4\}$
and 1$(M^J)^2 =
\pm I_4$, $tr M^J = 0$, $tr M^{I\dagger} M^{J} = 4\delta^{IJ}$.
See Appendix B.
In general, the evaluation of amplitudes in the BRST formalism is
carried out by the calculation of standard correlation functions, whose
singularity structure is set
from the operator product expansions\ref{10,11}.
For example, from (4.2) we find
$$\eqalignno{
\langle 0|S_{\dot\gamma} (z_1)\psi^\mu (z_2) S_\alpha (z_3) |0\rangle
&= \gamma^\mu_{\dot\gamma\alpha}{\scriptstyle{1\over\sqrt 2}}
(z_1 - z_2)^{-\half} (z_2 - z_3)^{-\half}\cr
\langle 0|\Sigma_{\dot n} (z_1) \Sigma_\ell (z_3) |0\rangle
&= (z_1 - z_3)^{-{3\over 4}} C^{-1}_{\dot n\ell}
\,.&(4.16)\cr}$$
%We can show
%$$tr M^J\alpha^a M^{I\dagger} = 8 f_{IaJ}\,\eqno(2.29)$$ where
%$f_{IaJ}$ is the 16-dimensional representation
%$2+2+2+2+2+2+2+2$ of $SU(2)$.

\vskip15pt
\centerline{\bf 5. Kinematics of the three-point amplitude}
\vskip 5pt
In order to identify on-shell dual model amplitudes with conventional
with conventional field theory couplings, we routinely use $k_i^2$ for massless
particles, which together with momentum conservation in the three-point
amplitudes, $k_1^\mu + k_2^\mu + k_3^\mu = 0$, implies
$$k_i\dt k_j = 0\,.\eqno(5.1)$$
In general, the dual model picks up a
particular gauge\ref{7,8} to describe the
graviton or gauge boson.
For gravitons, the harmonic condition satisfied in any
gauge
$$k^\mu\epsilon_{\mu\nu}=\hhalf k_\nu\epsilon^\mu_{\hskip3pt\mu}\eqno(5.2)$$
arises in the dual amplitude in the harmonic gauge:
$$\epsilon^\mu_{\hskip3pt\mu} =0\eqno(5.3a)$$
so that
$$k^\mu\epsilon_{\mu\nu} = 0\,.\eqno(5.3b)$$
Similarly, for gauge bosons, the string amplitudes pick up the Lorentz gauges
described by (3.6), so that we use also
$$k^\mu\epsilon_\mu^\pm = 0\,.\eqno(5.4)$$
In fact, further identities occur that are particularly useful in
identifying amplitudes involving states in the Ramond-Ramond sector as standard
field theory couplings\ref{6}. For the kinematics of the three-gluon
amplitude we have
$$\eqalignno{
&\epsilon^+_1\dt\epsilon^+_2 \epsilon^+_3\dt k_1
+ \epsilon^+_2\dt\epsilon^+_3 \epsilon^+_1\dt k_2
+ \epsilon^+_3\dt\epsilon^+_1 \epsilon^+_2\dt k_3 = 0&(5.5a)\cr
&\epsilon^-_1\dt\epsilon^-_2 \epsilon^-_3\dt k_1
+ \epsilon^-_2\dt\epsilon^-_3 \epsilon^-_1\dt k_2
+ \epsilon^-_3\dt\epsilon^-_1 \epsilon^-_2\dt k_3 = 0&(5.5b)\cr}$$
when all the polarizations are the same, even for complex momenta.
(For real momenta, the three-gluon amplitude vanishes for any choice of
polarizations\ref{6,12}, so one should be careful not to require real momenta,
which is no problem due to analyticity.)
For two polarizations of one helicity,
and the third polarization with opposite helicity, then
$$\epsilon^+_1\dt\epsilon^+_3  \epsilon^-_2\dt k_2 =0\eqno(5.5c)$$
and $$\epsilon^-_2\dt k_1 \epsilon^+_1\dt k_2 =0\eqno(5.5d)$$
so that
$$\eqalignno{
&\epsilon^+_1\dt\epsilon^-_2 \epsilon^+_3\dt k_1
+ \epsilon^-_2\dt\epsilon^+_3 \epsilon^+_1\dt k_2
+ \epsilon^+_3\dt\epsilon^+_1 \epsilon^-_2\dt k_3 =
\epsilon^+_1\dt\epsilon^-_2 \epsilon^+_3\dt k_1
+ \epsilon^-_2\dt\epsilon^+_3 \epsilon^+_1\dt k_2
&(5.5e)\cr}$$
with similar formulas for the other helicities.
The proof of (5.5) is as follows:
Consider
$$\eqalignno{
&i\epsilon^{\rho\kappa\mu\lambda}\epsilon^+_{1\rho}
k_{2\kappa}\epsilon^\pm_{2\mu}
\epsilon^+_{3\lambda}\cr
=&\pm (\epsilon^\pm_2\dt\epsilon^+_3 \epsilon^+_1\dt k_2
+ \epsilon^+_1\dt\epsilon^\pm_2 \epsilon^+_3\dt k_1 )&(5.6a)\cr
=&-i\epsilon^{\rho\kappa\mu\lambda}\epsilon^+_{1\rho}
(k_{1\kappa} + k_{3\kappa})\,
\epsilon^\pm_{2\mu} \epsilon^+_{3\lambda}\cr
=& - (\epsilon^+_3\dt\epsilon^+_1 \epsilon^\pm_2\dt k_3 +
\epsilon^+_1\dt\epsilon^\pm_2 \epsilon^+_3\dt k_1 )
-(\epsilon^\pm_2\dt\epsilon^+_3 \epsilon^+_1\dt k_2  +
\epsilon^+_3\dt\epsilon^+_1 \epsilon^\pm_2\dt k_3 )&(5.6b)\cr}$$
where (5.6a,b) are derived using (3.6).
Equating (5.6a) with (5.6b) we find (5.5a,c). Replacing $\epsilon_3$ with
$k_3$ in (5.6), we also find
$$\epsilon_2^-\dt k_3 \epsilon_1^+\dt k_3 = 0\eqno(5.7)$$ which is
(5.6d) using (5.4). Equations (5.5) are invariant under gauge transformations
$\epsilon_\mu\rightarrow k_\mu$ using (5.1).

We note that amplitudes involving states in the Ramond sector are
automatically zero for
certain combinations of polarizations. This is contrasted with
Neveu-Schwarz amplitudes
which can be calculated initially independent of the choice of polarizations.
A subsequent closer look at the Lorentz gauge condition (3.6) then reduces the
kinematics of the Neveu-Schwarz amplitudes to those of the
Ramond amplitudes for identical couplings. (See however the remark below
(5.14)).
{}From (4.14), we have the coupling of the general graviton tensor
with two massless Neveu-Schwarz vector mesons.
$$\eqalignno{
&\langle 0|V^a_{-1}(k_1,z_1)\,V_0(k_2,z_2,\epsilon_2)\,
V^b_{-1}(k_3,z_3) c(z_1)c(z_2)c(z_3)|0\rangle \cr
&\dt\langle 0|V_{-1}(k_1,\bar z_1,\epsilon_1)\,V_0(k_2,\bar z_2,\epsilon_2)\,
V_{-1}(k_3,\bar z_3,\epsilon_3) c(\bar z_1)c(\bar z_2)c(\bar z_3)|0\rangle \cr
&= \delta^{ab} \epsilon_2^{\mu\nu} k_{3\mu} [k_{3\nu} \epsilon_1\dt\epsilon_3
+ \epsilon_{3\nu} \epsilon_1\dt k_2 + \epsilon_{1\nu} \epsilon_3\dt k_1]\,.
&(5.8)\cr}$$
%When the polarizations $\epsilon_1^-, \epsilon_3^+$ as in (4.15),
%couple to the dilaton or to the antisymmetric tensor, the amplitude is zero.
For the polarizations $\epsilon_1^+, \epsilon_3^+$ coupling to the graviton
$h_{2\mu\nu}$, the amplitude (5.8) vanishes, as
we would expect from angular momentum conservation in the coupling of two spin
one mesons (with momenta of opposite sign) with a spin two particle.
% or a spin zero particle (either the
%dilaton or the anti-symmetric tensor).
The proof is as follows:
from (3.6) we find the identity
$$\eqalignno{
&(-\epsilon_3^{+\rho} k_3^\sigma +  k_3^\rho\epsilon_3^{+\sigma})
\eta^{\mu\nu}
+ (\epsilon_3^{+\rho} k_3^\nu -  k_3^\rho\epsilon_3^{+\nu})
\eta^{\mu\sigma}
+ (-\epsilon_3^{+\sigma} k_3^\nu +  k_3^\sigma\epsilon_3^{+\nu})
\eta^{\mu\rho} \cr
&= i (\epsilon_3^{+\mu} k_{3\omega} - k_3^\mu \epsilon_{3\omega}^+ )
\epsilon^{\omega\rho\sigma\nu}&(5.9)\cr}$$
Multiplying (5.9) by $\epsilon_{1\rho}^+ k_{1\sigma}$ we find
$$\eqalignno{&k_1^\mu\epsilon_3^{+\nu} \epsilon_1^+\dt k_2
+ k_3^\mu k_1^\nu \epsilon^+_1\dt\epsilon_3^+
+ k_3^\mu\epsilon_1^\nu\epsilon_3\dt k_2\cr
&= - k_1^\nu\epsilon_3^{+\mu} \epsilon_1^+\dt k_2
- k_3^\nu k_1^\mu \epsilon^+_1\dt\epsilon_3^+
- k_3^\nu\epsilon_1^\mu\epsilon_3\dt k_2
-\eta^{\mu\nu} \epsilon_1^+\dt k_3 \epsilon_3^+\dt k_1&(5.10)\cr}$$
{}From (5.10) and (5.3b) we find
$$\eqalignno{&h_2^{\mu\nu} k_{3\mu} [k_{3\nu}
\epsilon^+_1\dt\epsilon^+_3
+ \epsilon^+_{3\nu} \epsilon^+_1\dt k_2 +
\epsilon^+_{1\nu} \epsilon^+_3\dt k_1] = 0\,.&(5.11)\cr}$$
In the case of vector mesons from the Ramond-Ramond sector, the corresponding
amplitude vanishes identically. See (5.14).

For the polarizations $\epsilon_1^+, \epsilon_3^+$ coupling to the dilaton,
we use (3.18) and (5.5) to show that the amplitude (5.8) is given by
the non-zero expression
$$\eqalignno{
&\delta^{ab} D_2^{\mu\nu} k_{3\mu} [k_{3\nu} \epsilon^+_1\dt\epsilon^+_3
+ \epsilon^+_{3\nu} \epsilon^+_1\dt k_2 +
\epsilon^+_{1\nu} \epsilon^+_3\dt k_1]\cr
=& {\scriptstyle{1\over\sqrt 2}}
\epsilon_1^+\dt k_3 \epsilon_3^+\dt k_1&(5.12a)\cr
=& {\scriptstyle{2\over\sqrt 2}} \epsilon^+_{2\mu}\epsilon^-_{2\nu} k_{3\mu}
[\epsilon_{3\nu}^+\epsilon_1^+\dt k_2
+ \epsilon^+_{1\nu}\epsilon_3^+\dt k_1]\,.
&(5.12b)\cr}$$
The equality of (5.12a,b) follows from
$$\eqalignno{
&2\epsilon^+_2\dt k_3 [ \epsilon_2^-\dt\epsilon_3^+\epsilon^+_1\dt k_2
+ \epsilon_2^-\dt\epsilon_1^+\epsilon_3^+\dt k_1 ]\cr
= &2\epsilon^+_2\dt k_3 [ i k_{2\mu}\epsilon^+_{1\sigma}\epsilon^-_{2\kappa}
\epsilon^+_{3\lambda}\epsilon^{\mu\sigma\kappa\lambda} ] \cr
= &2\epsilon^+_{2\rho}\epsilon^-_{2\kappa} k_{2\mu}\epsilon^+_{1\sigma}
\epsilon^+_{3\lambda} k_3^\rho \,i \epsilon^{\mu\sigma\kappa\lambda} \cr
= &2 [ \epsilon^+_{2\rho}\epsilon^-_{2\kappa} k_{2\mu}\epsilon^+_{1\sigma}
i\epsilon_\lambda^{\hskip3pt\rho\alpha\beta}\epsilon^+_{3\alpha} k_{3\beta}
i \epsilon^{\mu\sigma\kappa\lambda} \cr
&\hskip3pt + \epsilon^+_{2\rho}\epsilon^-_{2\kappa}
k_{2\mu}\epsilon^+_{1\sigma}
\epsilon_3^{+\rho} k_{3\lambda} i \epsilon^{\mu\sigma\kappa\lambda} ] \cr
= &2 [\epsilon_1^+\dt\epsilon_2^+\epsilon_3^+\dt k_2\epsilon^-_2\dt k_3
+ \epsilon_2^+\dt\epsilon_2^-\epsilon_3^+\dt k_1 \epsilon_1^+\dt k_3
+ \epsilon_2^+\dt\epsilon_3^+\epsilon_1^+\dt k_2 \epsilon_2^-\dt k_1 ]\cr
= &2 \,\epsilon_2^+\dt\epsilon_2^-\epsilon_3^+\dt k_1 \epsilon_1^+\dt k_3 \cr
= &\,\epsilon_3^+\dt k_1 \epsilon_1^+\dt k_3\,. &(5.12c)\cr}$$
Similarly, for the polarizations $\epsilon_1^+, \epsilon_3^+$ coupling to the
pseudoscalar described by the antisymmetric tensor, we have
$$\eqalignno{
&\delta^{ab} B_2^{\mu\nu} k_{3\mu} [k_{3\nu} \epsilon^+_1\dt\epsilon^+_3
+ \epsilon^+_{3\nu} \epsilon^+_1\dt k_2 +
\epsilon^+_{1\nu} \epsilon^+_3\dt k_1]\cr
=& {\scriptstyle{1\over 4}} \epsilon_1^+\dt k_3 \epsilon_3^+\dt k_1&(5.13a)\cr
=& \hhalf\epsilon^+_{2\mu}\epsilon^-_{2\nu} k_{3\mu}
[\epsilon_{3\nu}^+\epsilon_1^+\dt k_2 + \epsilon^+_{1\nu}\epsilon_3^+\dt
k_1]\,.
&(5.13b)\cr}$$
For Ramond-Ramond mesons, the amplitudes corresponding to (5.12,13) vanish:
$$\eqalignno{
&\langle 0|V^{(1)}_{-{1\over 2}}(k_1,z_1)\,V_{-1}(k_2,z_2,\epsilon_2)\,
V^{(1)}_{-{1\over 2}}(k_3,z_3) c(z_1)c(z_2)c(z_3)|0\rangle \cr
&\dt\langle 0|V^{(1)}_{-{1\over 2}}(k_1,\bar z_1)\,V_{-1}(k_2,\bar
z_2,\epsilon_2)\,
V^{(1)}_{-{1\over 2}}(k_3,\bar z_3) c(\bar z_1)c(\bar z_2)c(\bar z_3)
|0\rangle \cr
& = u^{2\delta} (k_1) u^{2\gamma} (k_1)\epsilon_{2\mu}
\gamma^\mu_{\gamma\alpha} u^{1\alpha} (k_3) u^{1\beta} (k_3)
\epsilon_{2\nu}(\gamma^\nu)^T_{\beta\delta}
\hhalf f^{\dot k} f^{\dot n} C^{-1}_{\dot n\ell}
f^\ell f^m C^{-1}_{m\dot k}\cr
&= 0\,.
&(5.14)\cr}$$
since only $\gamma^\mu_{\dot\gamma\alpha}$ and $\gamma^\mu_{\gamma\dot\alpha}$
are non-zero. From (5.12,13,14) we see the Ramond mesons couple differently
to the dilaton (and to the antisymmetric tensor)
from the Neveu-Schwarz mesons. Thus in the superstring model of Ref. [13],
with $D=4$, $N=8$ supergravity that has 28 abelian massless vector mesons,
we find 12 of these couple to the dilaton as in (5.12) and the other 16 have
zero tree level coupling to the dilaton.
This feature already appears in $D=10$, $N=2$ supergravity\ref{14}, and
is useful in recent non-perturbative treatments of the superstring\ref{24}.
\vfill\eject

Using the four-dimensional covariant notation developed in sect. 3, we also
give the tree amplitude for the
{\it general graviton tensor coupling to two fermions}:
$$\eqalignno{
&\langle 0|V^{(2)}_{-{1\over 2}}(k_1,z_1)\,V_{-1}(k_2,z_2,\epsilon_2)\,
V^{(1)}_{-{1\over 2}}(k_3,z_3) c(z_1)c(z_2)c(z_3)|0\rangle \cr
&\dt\langle 0|V_{-1}(k_1,\bar z_1,\epsilon_1)\,V_0(k_2,\bar z_2,\epsilon_2)\,
V_{-1}(k_3,\bar z_3,\epsilon_3) c(\bar z_1)c(\bar z_2)c(\bar z_3)|0\rangle \cr
&= u^{2\dot\gamma}
(k_1)\epsilon_{2\mu} (-i\sigma^2\sigma^\mu)_{\dot\gamma\alpha}
u^{1\alpha} (k_3)\cr
&\hskip20pt\dt\epsilon_{2\nu}\, [k_3^\nu\epsilon_1\dt\epsilon_3
+ \epsilon^\nu_3 \epsilon_1\dt k_2 + \epsilon^\nu_1 \epsilon_3\dt k_1]\,
f^{\dot n} f^\ell C^{-1}_{\dot n\ell}\,.&(5.15)\cr}$$
(5.15) describes many processes. In the following the
extended supersymmetry indices $\dot n, \ell$ are suppressed.
Using the three-point on-shell kinematics from
sect. 5, we find for example, the one graviton -- two gravitino string tree
amplitudes:
$$\eqalignno{
&\epsilon^+_{2\mu}\epsilon^+_{2\nu}
u^{2\dot\gamma} (k_1)\epsilon_{2\mu} (-i\sigma^2\sigma^\mu)_{\dot\gamma\alpha}
u^{1\alpha} (k_3) [k_3^\nu\epsilon^-_1\dt\epsilon^+_3
+ \epsilon^{+\nu}_3 \epsilon^-_1\dt k_2
+ \epsilon^{-\nu}_1 \epsilon^+_3\dt k_1]\cr
= & h^+_{2\mu\nu}
\psi^{-\dot\gamma}_\rho (k_1) (-i\sigma^2\sigma^\mu)_{\dot\gamma\alpha}
\psi_\lambda^{+\alpha}(k_3)\, [k_3^\nu \eta^{\rho\lambda} +
k_1^\lambda\eta^{\nu\rho} ]
&(5.16a)\cr}$$
and
$$\eqalignno{
&\epsilon^-_{2\mu}\epsilon^-_{2\nu}
u^{2\dot\gamma} (k_1)\epsilon_{2\mu} (-i\sigma^2\sigma^\mu)_{\dot\gamma\alpha}
u^{1\alpha} (k_3) [k_3^\nu\epsilon^-_1\dt\epsilon^+_3
+ \epsilon^{+\nu}_3 \epsilon^-_1\dt k_2
+ \epsilon^{-\nu}_1 \epsilon^+_3\dt k_1]\cr
= & h^-_{2\mu\nu} \psi^{-\dot\gamma}_\rho (k_1)
(-i\sigma^2\sigma^\mu)_{\dot\gamma\alpha}
\psi_\lambda^{+\alpha}(k_3)\, [k_3^\nu \eta^{\rho\lambda} +
k_2^\rho\eta^{\lambda\nu} ]\,.
&(5.16b)\cr}$$
Note in (5.16) that the specific non-zero
helicity couplings appear automatically.
Also from (5.15), we find the one graviton -- two spin one-half fermion
amplitudes
$$\eqalignno{
&u^{2\dot\gamma} (k_1)\epsilon^+_{2\mu}
(-i\sigma^2\sigma^\mu)_{\dot\gamma\alpha}
u^{1\alpha} (k_3)\, \epsilon^+_{2\nu}\, [k_3^\nu\epsilon^+_1\dt\epsilon^-_3
+ \epsilon^{-\nu}_3 \epsilon^+_1\dt k_2
+ \epsilon^{+\nu}_1 \epsilon^-_3\dt k_1]\,\cr
&\hskip5pt
= h^+_{2\mu\nu} \chi^{+\dot\gamma}_\rho (k_1)
(-i\sigma^2\sigma^\mu)_{\dot\gamma\alpha}
\chi_\lambda^{-\alpha}(k_3)\,
[k_3^\nu \eta^{\rho\lambda} + k_2^\rho\eta^{\lambda\nu}]
\,;&(5.17a)\cr
&u^{2\dot\gamma} (k_1)\epsilon^-_{2\mu}
(-i\sigma^2\sigma^\mu)_{\dot\gamma\alpha}
u^{1\alpha} (k_3)\, \epsilon^-_{2\nu}\, [k_3^\nu\epsilon^+_1\dt\epsilon^-_3
+ \epsilon^{-\nu}_3 \epsilon^+_1\dt k_2
+ \epsilon^{+\nu}_1 \epsilon^-_3\dt k_1]\,\cr
&\hskip5pt
= h^-_{2\mu\nu} \chi^{+\dot\gamma}_\rho (k_1)
(-i\sigma^2\sigma^\mu)_{\dot\gamma\alpha}
\chi_\lambda^{-\alpha}(k_3)\, [k_3^\nu \eta^{\rho\lambda}
+ k_1^\lambda\eta^{\nu\rho} ]
\,.&(5.17b)\cr}$$
For spinors with parallel polarizations, the amplitude corresponding to (5.15)
vanishes as in (5.14):
$$\eqalignno{
&\langle 0|V^{(1)}_{-{1\over 2}}(k_1,z_1)\,V_{-1}(k_2,z_2,\epsilon_2)\,
V^{(1)}_{-{1\over 2}}(k_3,z_3) c(z_1)c(z_2)c(z_3)|0\rangle \cr
&\dt\langle 0|V_{-1}(k_1,\bar z_1,\epsilon_1)\,V_0(k_2,\bar z_2,\epsilon_2)\,
V_{-1}(k_3,\bar z_3,\epsilon_3) c(\bar z_1)c(\bar z_2)c(\bar z_3)|0\rangle \cr
&=0\,. &(5.18)\cr}$$
(5.15,18) show for example the graviton makes no transition between
spin ${\scriptstyle{3\over 2}}$ and spin $\hhalf$,
since from (5.15),
$u^{2\dot\gamma} (k_1)\epsilon^+_{2\mu}
(-i\sigma^2\sigma^\mu)_{\dot\gamma\alpha}
u^{1\alpha} (k_3)
\epsilon^+_{2\nu}\, {[k_3^\nu\epsilon^+_1\dt\epsilon^+_3
+ \epsilon^{+\nu}_3 \epsilon^+_1\dt k_2 +
\epsilon^{+\nu}_1 \epsilon^+_3\dt k_1]\nobreak=0}$,
and from (5.18),
$u^{1\gamma} (k_1)\epsilon^+_{2\mu}\gamma^\mu_{\gamma\alpha}
u^{1\alpha} (k_3)\,
\epsilon^+_{2\nu}\, [k_3^\nu\epsilon^-_1\dt\epsilon^+_3
+ \epsilon^{+\nu}_3 \epsilon^-_1\dt k_2
+ \epsilon^{-\nu}_1 \epsilon^+_3\dt k_1] = 0$,
etc. Amplitudes for the remaining
combinations of spinor polarizations, including spinors for the left instead of
right-movers also occur and are similar to (5.15,17). Additional spin $\hhalf$
fermions also appear from left-moving scalars and right-moving spinors, and
visa versa.
\vskip15pt
\centerline{\bf 6. Non-abelian coupling}
\vskip 5pt
%We consider the vertex operator directly in four dimensions.
%$$\eqalignno{V^{(1)}_{-{3\over 2}}(k,z) &= v^{1{\dot\alpha}}(k)
%S_{\dot\alpha}(z) e^{ik\dt X(z)} f^\ell
%\Sigma_\ell (z)
%\,V(z)\,e^{-{3\over 2}\phi(z)}&(6.1a)\cr
%V^{(2)}_{-{3\over 2}}(k,z) &= v^{2{\alpha}}(k)
%S_{\alpha}(z) e^{ik\dt X(z)}
%f^{\dot\ell}\Sigma_{\dot\ell}(z)
%\,V(z)\,e^{-{3\over 2}\phi(z)}\,.&(6.1b)\cr}$$
In this section, we discuss a set of tree amplitudes involving
massless gauge boson states coming from the Ramond-Ramond sector, which
exhibit a non-abelian structure. In the spirit of previous
investigations [15], the
gauge symmetry is enlarged by modifying the choice of the worldsheet
supercurrent.
Although such states are forbidden
by the conventional analysis of [16],
nonetheless we find their kinematic structure interesting, and
suggest one possible mechanism by which they could be incorporated into
an interacting string model satisfying space-time unitarity.
This mechanism involves a non-hermitian piece of the internal supercurrent
$\tilde F(z)$.
The supercurrent is given by
$$F(z) = a_\mu(z) \psi^\mu(z) + \bar F(z)\,\eqno(6.1)$$
where
$0\le\mu\le 3$ and we choose
$$\eqalignno{\bar F(z) = &\tilde\gamma^5
\otimes
\widehat F (z)
+\,\,\tilde\gamma^5\otimes\tilde\Gamma^7\otimes\,\tilde F(z)\,&(6.2)\cr}$$
where
$\widehat F (z)\equiv
(-{\textstyle{i\over 6}})
{\textstyle{1\over{\scriptstyle{ \sqrt{c_\psi\over2}}}}}
f_{abc} \psi^a(z) \psi^b(z) \psi^c(z)\,$ and
$f_{abc}$ given by the totally antisymmetric structure constants of
$SU(2)\otimes SU(2)$ with $f_{abc}f_{abe} = c_\psi\delta_{ce}$.
%$$\eqalignno{(-{\textstyle{i\over 6}})
%{\scriptstyle{1\over{\scriptstyle{ \sqrt{c_\psi\over2}}}}}
%f_{abc} \psi^a(z) \psi^b(z) \psi^c(z)\,\Sigma_\ell(\zeta)
%&= (z-\zeta)^{-{3\over 2}}
%(-{i\over{\scriptstyle 2{\sqrt 2}}})
%(1+i)\, (I_4)_\ell^{\hskip3pt{\dot\ell}}\Sigma_{\dot\ell}(\zeta) +\ldots\cr
%(-{\textstyle{i\over 6}})
%{\scriptstyle{1\over{\scriptstyle{ \sqrt{c_\psi\over2}}}}}
%f_{abc} \psi^a(z) \psi^b(z) \psi^c(z)\,\Sigma_{\dot\ell}(\zeta)
%&= (z-\zeta)^{-{3\over 2}}
%(-{i\over{\scriptstyle2{\sqrt 2}}})
%(-1+i)\, (I_4)_{\dot\ell}^{\hskip3pt{\ell}}\Sigma_{\ell}(\zeta) +\ldots\,\cr
%&&(6.3)\cr}$$
%where we use operator products of the form
%$$f_{abc} \psi^a(z) \psi^b(z) \psi^c(z)\,\Sigma_\ell(\zeta) =
%(z-\zeta)^{-{3\over 2}} f_{abc} {1\over{\scriptstyle 2{\sqrt 2}}}
%(\Gamma^a\Gamma^b\Gamma^c)_\ell^{\hskip3pt\dot\ell}
%\Sigma_{\dot\ell}(\zeta)\,+\ldots\,.\eqno(6.4)$$
For ${\cal S}(z) \equiv f^\ell\Sigma_\ell (z)$ and
$\tilde{\cal S}(z)\equiv \hhalf e^{-{i\pi\over 4}} f^{\dot\ell}
\Sigma_{\dot\ell} (z)$, we have
$$\eqalignno{
\widehat F (z) \,{\cal S}(\zeta)
&= (z-\zeta)^{-{3\over 2}}\,
\tilde{\cal S}(\zeta) +\ldots&(6.3)\cr
\widehat F(z) \,\tilde{\cal S}(\zeta)
&= (z-\zeta)^{-{3\over 2}}\,
{\textstyle{1\over 4}} {\cal S}(\zeta) +\ldots
&(6.4)\cr}$$
where we use operator products of the form
$$f_{abc} \psi^a(z) \psi^b(z) \psi^c(z)\,\Sigma_\ell(\zeta) =
(z-\zeta)^{-{3\over 2}} f_{abc} {1\over{\scriptstyle 2{\sqrt 2}}}
(\Gamma^a\Gamma^b\Gamma^c)_\ell^{\hskip3pt\dot\ell}
\Sigma_{\dot\ell}(\zeta)\,+\ldots\,.\eqno(6.5)$$
$\tilde F(z)$ corresponds to the remaining
internal degrees of freedom with $c=6$. We assume there exists an
anti-periodic supercurrent, and a pair of
states in the
Ramond sector of this piece of the internal system corresponding to
weight zero conformal spin fields $V(z)$ and $U(z)$ such that
$$\eqalignno
{\tilde F(z) V(\zeta) &= (z-\zeta)^{-{3\over 2}} U(\zeta) +\ldots\cr
\tilde F(z) U(\zeta) &= (z-\zeta)^{-{3\over 2}} (-{\textstyle{c\over 24}})
V(\zeta) + \ldots\,.&
(6.5)\cr}$$
(6.5) corresponds to a non-hermitian choice for the operator $\tilde F_0$,
that is to say non-hermitian with respect to a vector space of
states with positive-definite inner product.
This is because in such a space, the eigenvalues of the square of a
hermitian operator are non-negative, but from (6.5) we see
$\tilde F_0^2 V(0) |0\rangle = -{\scriptstyle{1\over 4}}  V(0) |0\rangle $,
i.e., a negative eigenvalue.
Note that the
definition of the
hermitian adjoint for any operator $A$, which is
$$(A\psi_a,\psi_b) = (\psi_a, A^\dagger\psi_b)\,,\eqno(6.6a)$$
depends on the definition of the inner product
$$(\psi_a,\psi_b)\,,\eqno(6.6b)$$
since (6.6b) is used to evaluate (6.6a).
One example of a conformal field theory
where $F_0^2$ can take on negative eigenvalues is
the $b,c,\beta,\gamma$ superconformal ghost system; here $F_0$ is not
hermitian with respect to a non-vanishing inner product, given for
example by the adjoint state defined as
$( c_1|0\rangle)^\dagger = \langle 0| c_{-1} c_0 $.
(Although $F_0$ is hermitian with respect to a vanishing norm, given for
example by the adjoint state defined as
$( c_1|0\rangle)^\dagger = \langle 0| c_{-1}$.)
Another example is the combined system $(T_m, (b_1, c_1))$ with
$N=1$ superconformal symmetry with $c=15$ of Berkovits and Vafa\ref{17}
representing the matter system of an $N=1$ fermionic string whose states
physical states are in one-to-one correspondence with the states of
the bosonic string with $c=26$.

We also remark that
since the norm of a state $\tilde F_0 |\Psi\rangle$ is given by
$||\tilde F_0 |\Psi\rangle || = \langle \Psi | \tilde F_0^\dagger
\tilde F_0 |\Psi\rangle$, that therefore
non-hermitian $\tilde F_0$ does not imply $\tilde F_0 |\Psi\rangle$ is
a negative norm state, even though
$\langle\Psi|\tilde F_0^2 |\Psi\rangle < 0$.

The states of the Ramond sector are created by conformal fields called
spin fields\ref{18-22}. The spin fields are nonlocal with
respect to the worldsheet supercurrent $T_F$ because
they make states in the Ramond sector of the theory,
whereas superconformal fields corresponding to states in the Neveu-Schwarz
sector are local with respect to $T_F$.
In general, the operator product of a spin field with
any fermionic part of a superfield is nonlocal (i.e. double-valued)
since spin fields change the boundary condition on fermion fields
between periodic and anti-periodic.
%The OPE of a spin
%field with the fermionic parts of the NS superfields is nonlocal (i.e.
%double-valued) in order to generate the correct Fourier series expansion
%of fermions in the Ramond sector; spin fields flip the boundary
%condition on the fermion fields between periodic and anti-periodic.
%A spin field may be represented as the endpoint of a branch cut in the
%fermion fields.

In the Ramond sector, the operator products of the
spin fields with $T_F$, have a $(z-\zeta)^{-{3\over2}}$ singularity;
all except for the ``ground states'',
(i.e. those for which $h={c\over 24}$), which have only a
$(z-\zeta)^{-{1\over 2}}$ singularity with $T_F$. Since $L_0$ and $F_0$
commute,
all excited states are in pairs $S^\pm (z) |0\rangle$ related by
$F_0$, only the ``ground states'' need not be paired.
We have
$$\eqalignno{|h^+\rangle &= S^+(0) |0\rangle\cr
|h^-\rangle &= S^-(0) |0\rangle = F_0 |h^+\rangle\cr
(h -{\textstyle{c\over 24}})|h^+\rangle &= F_0 |h^-\rangle\cr
L_0 |h^\pm\rangle &= h |h^\pm\rangle\cr}$$
or in terms of the fields:
$$\eqalignno{F(z) S^+(\zeta) &= (z-\zeta)^{-{3\over2}} S^-(\zeta)\cr
F(z) S^-(\zeta) &= [h-{\textstyle{c\over 24}}]
\,(z-\zeta)^{-{3\over2}} S^+(\zeta)\,.\cr}$$
If $ h = {c\over 24}$, global worldsheet supersymmetry is unbroken in the
Ramond sector; where $F_0$ is the global supersymmetry charge satisfying
the global supersymmetry algebra
$F^2_0 = L_0 - {c\over 24}$.
In this case the states need not come in pairs, i.e. only
$|h^-\rangle$ survives since
if $F_0$ is hermitian, then for $h={\textstyle{c\over 24}}$ we have
$\langle h^- | h^-\rangle = \langle h^+ | F^2_0 |h^+\rangle =
\langle h^+ | L_0 - {\textstyle{c\over 24}} | h^+\rangle = 0$, i.e.
$|h^-\rangle$ is null, i.e. has zero norm.

The operator product expansions for the pair of spin fields $U(z), V(z)$
are chosen to be
$$\eqalignno{
V(z) V(\zeta)&= (z-\zeta)^0 C_0(\zeta) +\dots\cr
U(z) U(\zeta)&= (z-\zeta)^0 C'_0(\zeta) +\ldots\cr
V(z) U(\zeta)&= \ldots + (z-\zeta)^{\textstyle {-{1\over 2}}} C_{-{1\over 2}}
(\zeta) + \ldots\cr
U(z) V(\zeta)&= \ldots - (z-\zeta)^{\textstyle {-{1\over 2}}} C_{-{1\over 2}}
(\zeta) +\ldots
\,&(6.7)\cr}$$
where the subleading terms are less singular by integer powers of
$(z-\zeta)$, and for example
$C_{-{1\over 2}}(z)$ is some dimension ${\textstyle{-{1\over 2}}}$
operator whose coefficient may be
zero apriori, and which is understood to represent $\tilde\gamma^5\otimes
\tilde\Gamma^7\otimes C_{-{1\over 2}}(z)$; and
$C_0 (z)$ has weight zero, etc.
(6.7) together with (4.2) are consistent with locality of the string vertices
given in (6.1).

For $\Sigma^{(1)}(z) \equiv {\cal S}(z) V(z)$ and
$\tilde\Sigma^{(1)}(z)\equiv
\tilde{\cal S}(z) V(z) + {\cal S}(z) U(z)$, BRST invariant
vertex operators for the Ramond states in the canonical $q=-\half$
superconformal ghost picture are given for $u,v$ in (3.2),(3.3) by
%$$\eqalignno{V^{(1)}_{-\half}(k,\zeta)
%&=\lim_{z\rightarrow\zeta} e^{\phi(z)} F(z)
%V^{(1)}_{-{3\over 2}}(k,\zeta)\cr
%&=[u^{1\alpha} (k) S_\alpha (\zeta) e^{ik\dt X(\zeta)} f^\ell \Sigma_\ell
%(\zeta) V(\zeta)\cr
%&\hskip 8pt -
%v^{1\dot\alpha}(k) S_{\dot\alpha} (\zeta) e^{ik\dt X(\zeta)}
%(-{\textstyle{i\over {2\sqrt 2}}}) (1+i)f^{\dot\ell}
%\Sigma_{\dot\ell}(\zeta)
%V(\zeta)\cr
%&\hskip 8pt -
%v^{1\dot\alpha}(k) S_{\dot\alpha} (\zeta) e^{ik\dt X(\zeta)}\,
%f^\ell \Sigma_\ell (\zeta)\,U(\zeta) ]
%e^{-\half\phi(\zeta)}\,.&(6.8a)\cr}$$
%$$\eqalignno{V^{(2)}_{-\half}(k,\zeta)
%& = \lim_{z\rightarrow\zeta} e^{\phi(z)} F(z)
%V^{(2)}_{-{3\over 2}}(k,\zeta)\cr
%&=[-u^{2\dot\alpha} (k) S_{\dot\alpha}
%(\zeta) e^{ik\dt X(\zeta)}
%f^{\dot\ell}\Sigma_{\dot\ell}(\zeta))\,V(\zeta)\cr
%&\hskip 12pt +
%v^{2\alpha}(k) S_{\alpha} (\zeta) e^{ik\dt X(\zeta)}
%(-{\textstyle{i\over {2\sqrt 2}}})
%(-1+i) f^\ell\Sigma_\ell(\zeta) V(\zeta)\cr
%&\hskip 12pt -
%v^{2\alpha}(k) S_{\alpha} (\zeta) e^{ik\dt X(\zeta)}\,
%f^{\dot\ell}\Sigma_{\dot\ell}(\zeta)\,
%U(\zeta)\,]\,
%e^{-\half\phi(\zeta)}\,.&(6.8b)\cr}$$
$$\eqalignno{V^{(1)}_{-\half}(k,\zeta)
&=
%A\,
[\, u^{1\alpha} (k) S_\alpha (\zeta) \,\Sigma^{(1)}(\zeta)
- v^{1\dot\alpha}(k) S_{\dot\alpha} (\zeta)
\,\tilde \Sigma^{(1)}(\zeta)\,]\,
\,e^{ik\dt X(\zeta)}\,
e^{-\half\phi(\zeta)}\,,&(6.8a)\cr}$$
and for $\Sigma^{(2)}(z) \equiv\tilde{\cal S}(z) V(z)$ and
$\tilde\Sigma^{(2)}(z)\equiv
{\textstyle{{1\over 4}}} {\cal S}(z) V(z) - \tilde{\cal S}(z) U(z)$,
%$$\eqalignno{V^{(2)}_{-\half}(k,\zeta)
%&=
%-2 e^{i\pi\over 4} [\,u^{2\dot\alpha} (k) S_{\dot\alpha} (\zeta)
%\,\tilde\Sigma (\zeta))\,V(\zeta)\cr
%&\hskip 35pt +
%v^{2\alpha}(k) S_{\alpha} (\zeta)
%[\,-{\textstyle{1\over 4}}\Sigma (\zeta) V(\zeta) + \tilde\Sigma (\zeta)\,
%U(\zeta)\,]\,]\,e^{ik\dt X(\zeta)}\,
%e^{-\half\phi(\zeta)}\,.&(6.8b)\cr}$$
$$\eqalignno{V^{(2)}_{-\half}(k,\zeta)
&=
%B
\, [\,u^{2\dot\alpha} (k) S_{\dot\alpha} (\zeta)
\Sigma^{(2)} (\zeta)
- v^{2\alpha}(k) S_{\alpha} (\zeta)\,\tilde\Sigma^{(2)}\,]
\,e^{ik\dt X(\zeta)}\,
e^{-\half\phi(\zeta)}\,.&(6.8b)\cr}$$
We note that within the fermion scattering amplitudes,
the modified fermion emission vertex given above effectively
normalizes the Ramond-Ramond bosons to $1$ not $k^0$, i.e. the same
normalization as the conventional gauge bosons found in the NS-NS sector,
a required feature
when both kinds of bosons are used together to form the adjoint representation
of a bigger group.
Consider the following amplitude.
On the left-moving side we have
$$\eqalignno{
&\langle 0|V^{(1)}_{-{1\over 2}}(k_1,z_1)\,V_{-1}(k_2,z_2,\epsilon_2)\,
V^{(1)}_{-{1\over 2}}(k_3,z_3) c(z_1)c(z_2)c(z_3)|0\rangle \cr
=&\langle 0| [u^{1\gamma} (k_1) S_\gamma(z_1) \Sigma^{(1)}(z_1)
- v^{1\dot\gamma}(k_1) S_{\dot\gamma} (z_1)
\tilde\Sigma^{(1)}(z_1) \,]
\,e^{ik_1\dt X(z_1)}\,
e^{-\half\phi(z_1)}\cr
&\hskip 10pt\dt [\epsilon_{2\mu} \psi^\mu (z_2)
\, e^{ik_2\dt X(z_2)} \,] e^{-\phi (z_2)}\cr
&\hskip 10pt\dt[u^{1\alpha} (k_3) S_\alpha (z_3)\Sigma^{(1)}(z_3)
- v^{1\dot\alpha}(k_3) S_{\dot\alpha} (z_3) \tilde\Sigma^{(1)}(z_3)\,]
\,e^{ik_3\dt X(z_3)}\,
e^{-\half\phi(z_3)} |0\rangle\cr
&\hskip 4pt\dt \langle 0|c(z_1)c(z_2)c(z_3)|0\rangle \cr
&= - [u^{1\gamma} (k_1) \epsilon_{2\mu}
\gamma^\mu_{\gamma\dot\alpha} v^{1\dot\alpha}(k_3)
+ v^{1\dot\gamma} (k_1) \epsilon_{2\mu}
\gamma^\mu_{\dot\gamma\alpha}u^{1\alpha}
(k_3)\,]\, {\textstyle{1\over\sqrt 2}}\cr
&\hskip 25pt\dt
f^n f^{\dot\ell}\, C^{-1}_{n\dot\ell}
\,\hhalf e^{-{i\pi\over 4}}\,
%A^2
&(6.9)\cr}$$
\vfill\eject
On the right-moving side we consider
$$\eqalignno{
&\langle 0|V^{(1)}_{-{1\over 2}}(k_1,z_1)\,V^a_{-1}(k_2,z_2)\,
V^{(1)}_{-{1\over 2}}(k_3,z_3) c(z_1)c(z_2)c(z_3)|0\rangle \cr
=&\langle 0|[u^{1\delta} (k_1) S_\delta(z_1) \Sigma^{(1)}(z_1)\cr
&\hskip 8pt -
v^{1\dot\delta}(k_1) S_{\dot\delta} (z_1)
\tilde\Sigma^{(1)}(z_1)\,]\,e^{ik_1\dt X(z_1)}\,
e^{-\half\phi(z_1)}\cr
&\hskip 10pt\dt [\tilde\gamma^5 \otimes \psi^a(z_2)
\, e^{ik_2\dt X(z_2)} \,] e^{-\phi (z_2)}\cr
&\hskip 10pt\dt[u^{1\beta} (k_3) S_\beta (z_3)\Sigma^{(1)}(z_3)\cr
&\hskip 8pt -
v^{1\dot\beta}(k_3) S_{\dot\beta} (z_3)
\tilde\Sigma^{(1)}(z_3)\,]\,e^{ik_3\dt X(z_3)}\,
e^{-\half\phi(z_3)} |0\rangle\cr
&\hskip 4pt\dt \langle 0|c(z_1)c(z_2)c(z_3)|0\rangle \cr
&= u^{1\delta} (k_1)
C^{-1}_{\delta\beta} u^{1\beta}(k_3) f^k f^m (-\alpha^a_{km})
{\textstyle{1\over\sqrt 2}}
%A^2
&(6.10)\cr}$$
where in (6.10) we have assumed $C'_0 = {\textstyle{i\over 4}} C_0$.
Combining the left and right-moving pieces (6.9) and (6.10) we find
up to an overall normalization constant
$$\eqalignno
{A_3&= i f^k f^n C^{-1}_{n\dot\ell} f^{\dot\ell} f^m \alpha^a_{mk}
{\textstyle{1\over 2}}
%e^{-{i\pi\over 4}}\,
%A^4
\cr
&\hskip8pt\dt
\epsilon^-_{2\mu}
[\,u^{1\delta}(k_1) u^{1\gamma} (k_1)
(-i\sigma^2\bar\sigma^\mu)_{\gamma\dot\alpha}
v^{1\dot\alpha} (k_3) u^{1\beta} (k_3) (i\sigma^2)_{\beta\delta}
{\textstyle{1\over\sqrt 2}}\cr
&\hskip20pt + u^{1\delta}(k_1) v^{1\dot\gamma} (k_1)
(-i\sigma^2\sigma^\mu)_{\dot\gamma\alpha}
u^{1\alpha} (k_3) u^{1\beta} (k_3) (i\sigma^2)_{\beta\delta}
{\textstyle{1\over\sqrt 2}}\,]\cr
&= i f^k f^n C^{-1}_{n\dot\ell} f^{\dot\ell} f^m \alpha^a_{mk}
{\textstyle{1\over 2}}
%e^{-{i\pi\over 4}}\,
%A^4
\cr
&\hskip8pt\dt
{\rm tr}
[\,(-\epsilon^+_{1\rho} k_{1\sigma}\bar\sigma^\rho\sigma^\sigma\sigma^2)
(-i\sigma^2\bar\sigma^\mu)(\epsilon_{3\lambda}^+\sigma^\lambda\sigma^2)
(i\sigma^2)\cr
&\hskip24pt +(\epsilon^+_{1\rho}\bar\sigma^\rho\sigma^2)(-i\sigma^2\sigma^\mu)
(-\epsilon_{3\lambda}^+ k_{3\kappa}\bar\sigma^\lambda\sigma^\kappa\sigma^2)
(i\sigma^2)\,]\cr
&= i f_{IaJ} \,
%{\textstyle{1\over{2\sqrt 2}}} e^{-{i\pi\over 4}}\, A^4\cr
[\,\epsilon^+_1\dt\epsilon_2^-\epsilon^+_3\dt k_1
+ \epsilon^-_2\dt\epsilon_3^+ \epsilon_1^+\dt k_2\,]
&(6.11)\cr}$$
which is the three-gluon tree coupling for this set of polarizations, as
$\epsilon^+_1\dt\epsilon_3^+ \epsilon_2^-\dt k_3 = 0$.
The structure constants
$f_{IaJ} = 2\,{\rm tr} (M_I^\dagger M_J\alpha^a)$ form the (2,2,2,2)
representation\ref{6}
of $SU(2)^4$ and suggest that the symmetry group is enhanced to $SO(8)$ from
its symmetric subgroup $SU(2)^4$.
Generalizations to other polarizations must be constructed consistent with
the required couplings\ref{23}. Also, couplings to other states such
as the graviton must be of the conventional form.
This can involve modifying the superconformal fields corresponding to the
Neveu-Schwarz states by the internal $\tilde F$ conformal theory as well
as the above modification of the spin fields.
For example, in (6.9) and (6.10) we can replace
$V_{-1}(k_2,z_2,\epsilon_2) = \epsilon\dt\psi e^{ik\dt X}$
and $V^a_{-1}(k_2,z_2) = \tilde\gamma\otimes\psi^a e^{ik\dt X}$
by forms similar to
$\epsilon\dt\psi C_0 + (\epsilon\dt a + k\dt\psi\epsilon\dt\psi)C_{-\half}$
and {$\tilde\gamma\otimes\psi^a C_0 +\nobreak(k\dt\psi\psi^a -\nobreak
{i\over 2}
f_{abc}\psi^b\psi^c) C_{-\half}$} respectively. Here $C_{-\half}$ and
$C_{0}$ form the lower and upper components of a superconformal field,
and the modified vertices remain BRST invariant.
In this way, the states in the hermitian part of the theory could be
in one-to-one correspondence with the physical states in the entire theory.
These modified vertices yield expressions in (6.11) with all
polarizations present.
\vskip15pt
\vfil\eject
\centerline{\bf 7. Modular invariance}
\vskip 5pt
%By conformal invariance, $\Lambda$ vanishes at the tree level for any
%string model. In superstring theory, $\Lambda$ vanishes at one-loop
%and it is conjectured this remains true to all orders.
%It is related to the function which counts the number of
%states at each mass level.
To describe consistent interacting strings, one must in general check that
all the scattering amplitudes are modular invariant, finite, and unitary.
A guide to this program is the calculation of the partition function,
i.e. the one-loop cosmological constant $\Lambda$,
which can be checked for modular invariance, albeit a quantity equal to zero.
For closed strings, the one-loop cosmological constant is defined by
$$\Lambda\equiv \hhalf {\rm tr\,ln} \Delta^{-1}\,\eqno(7.1)$$ where
$$\Delta^{-1} = \alpha' (p^2 + m^2)\,;\qquad
\hhalf\alpha' m^2 = \alpha' m_L^2 + \alpha' m_R^2 \,\eqno(7.2)$$
so in $D$ space-time dimensions, for $\omega = e^{2\pi i\tau}$,
$$\eqalignno{\Lambda &= -\hhalf (2\pi)^{-1}
(\alpha')^{-{D\over 2}}\int_F d^2\tau
(Im\tau)^{-2-\half ({D - 2})}\cr
&\hskip5pt\cdot\sum_{\rm all\, sectors}
{\rm tr} [\bar\omega^{\alpha' m_L^2} \omega^{\alpha' m_R^2}
{\rm possible\, projections}]\,.&(7.3)\cr}$$
$F$ is a fundamental region of the modular group:
$\half\le Re\tau\le\half\,;\,|\tau|> 1.$
The unitary $D=4$, $N=8$ superstring model considered in [13],
which incorporates
(4.2b) as {\it part}
of the internal field theory, has its one-loop modular invariant
partition function given by
$$\eqalignno{\Lambda &= -({4\pi{\alpha'}^2})^{-1}
\int_F d^2\tau (Im\tau)^{-3} |f(\omega)|^{-12} |\omega|^{-\half}\cr
&\hskip70pt\cdot {\textstyle{1\over 4}}[\theta_3^4 -\theta_4^4 -\theta_2^4]
[\bar\theta_3^4 -\bar\theta_4^4 -\bar\theta_2^4]\cr
&\hskip70pt\cdot [\,\hhalf ( |\theta_3 |^{12} + |\theta_4 |^{12} +
|\theta_2 |^{12} ) \, |f(\omega)|^{-12} |\omega|^{-\half}\,]\,. &(7.4)\cr}$$
For the $D=4$, $N=8$ model considered in sect. 6,
the partition function computed for
the degrees of freedom donoted in (4.2) by $a_\mu, \psi^\mu$ and $\psi^a$ with
$\alpha' m_L^2 = L_0 -{\textstyle c\over 24}$, etc. is
$$\eqalignno{\Lambda &= -({4\pi{\alpha'}^2})^{-1}
\int_F d^2\tau (Im\tau)^{-3} |f(\omega)|^{-12} |\omega|^{-\half}\cr
&\hskip70pt\cdot {\textstyle{1\over 4}}[\theta_3^4 -\theta_4^4 -\theta_2^4]
[\bar\theta_3^4 -\bar\theta_4^4 -\bar\theta_2^4]\,.&(7.5)\cr}$$
We see (7.5) is already modular invariant using the transformation properties
under $SL(2,I)$ given by
$$\eqalignno{\tau\rightarrow\tau + 1\,:\quad&\theta_3\rightarrow\theta_4;
\,\theta_4\rightarrow\theta_3;\,\theta_2\rightarrow e^{i{\pi\over 4}}
\theta_2\,;\,
{\rm Im}\tau\rightarrow {\rm Im}\tau\cr
\tau\rightarrow -{1\over\tau} :\quad&
\theta_2\rightarrow (-i\tau)^\half\theta_4;\, \theta_4\rightarrow
(-i\tau)^\half\theta_2;\,\theta_3\rightarrow (-i\tau)^\half\theta_3;\,
{\rm Im}\tau\rightarrow |\tau|^{-2} {\rm Im}\tau;\cr
&\omega^{1\over 24} f(\omega)\rightarrow (-i\tau)^\half
\omega^{1\over 24} f(\omega)&(7.6)\cr}$$
where
$f(\omega) = \prod_{n=1} (1 -\omega)$ is related to
the Dedekind eta function $\eta(\omega) = \omega^{1\over 24} f(\omega)$,
$\omega =\nobreak e^{2\pi i\tau}$
and $\theta_i(0|\tau)$ are the Jacobi theta functions.
We suggest that the inclusion of the remaining internal conformal field theory
will leave the modular invariance of (7.5) unchanged in the theory
postulated in sect.6.; and
the physical states may be
in one to one correspondence with the states of the partition function
given in (7.5). This latter
feature of adding conformal fields to the matter sector
without changing the spectrum has been discussed in a different context in
[17].
\vfill\eject
\centerline{\bf 8. Conclusions}
\vskip 5pt
We have developed a formalism for studying four-dimensional massless
spin fields, and computed their
BRST invariant tree amplitudes. In sect.'s 6 and 7,
the physical states in the $q=-\hhalf$ ghost picture, given by say (6.8a),
$$V^{(1)}_{-\half}(k,0) |0\rangle\eqno(8.1)$$
satisfy, for $F_0$ given in (6.2):
$$F_0 \,V^{(1)}_{-\half}(k,0) |0\rangle = 0\,,\eqno(8.2)$$ which is the
physical state condition in the old covariant formalism. We can either
check (8.2) explicitly, or recall that it follows from the BRST invariance
of (8.1). These states have the additional feature that
$$\eqalignno{F_0^{s.t.} V^{(1)}_{-\half}(k,0) |0\rangle&=
- u^{1\alpha} (k) S_\alpha (0) \tilde\Sigma (0) e^{ik\dt X(0)}
|0\rangle
\ne 0&(8.3a)\cr
\bar F_0\,V^{(1)}_{-\half}(k,0) |0\rangle&=
u^{1\alpha} (k) S_\alpha (0) \tilde\Sigma (0)\,e^{ik\dt X(0)}
|0\rangle
\ne 0\,,&(8.3b)\cr}$$
where $F^{s.t.} (z) \equiv a_\mu (z)\psi^\mu (z)$,
even though the sum $F_0 = F_0^{s.t.} + \bar F_0$ does annihiliate
the states as is
shown in (8.2).
Of course (8.3a) still describes massless states since although
$k\dt\gamma v = u\ne 0$, we have
$$-m^2 v = k^2 v = k\dt\gamma u = 0\,,\eqno(8.4)$$
so that
$$ (F_0^{s.t.})^2 \,V^{(1)}_{-\half}(k,0) |0\rangle = 0\,.\eqno(8.5)$$
Now, $F_0\,V^{(1)}_{-\half}(k,0) |0\rangle = 0$ implies
$F_0^2\,V^{(1)}_{-\half}(k,0) |0\rangle = 0$. Using
$\{ F_0^{s.t.} , \bar F_0\} = 0$ and (8.5), we find
$$F_0^2\, V^{(1)}_{-\half}(k,0) |0\rangle
= ((F_0^{s.t.})^2 + \bar F_0^2) \,V^{(1)}_{-\half}(k,0)
|0\rangle = 0\,,\eqno(8.6)$$
and thus
$$\bar F_0^2\,V^{(1)}_{-\half}(k,0) |0\rangle = 0\,.\eqno(8.7)$$
So we have
$\bar F_0^2\,V^{(1)}_{-\half}(k,0) |0\rangle = 0$, but from (8.3b) that
$\bar F_0\,V^{(1)}_{-\half}(k,0) |0\rangle \ne 0$. This is consistent with the
observation that $\tilde F_0$ which forms part of $\bar F_0$ is not hermitian.
(8.3b) indicates that global worldsheet supersymmetry generated by the
charge $\bar F_0$ is broken in the Ramond sector.
Consistency of the one-loop amplitudes is being investigated\ref{23}.

The mechanism we have suggested yields a string model with the gauge group
derived by enhancing the symmetric subgroup $SU(2)^4$ to $SO(8)$,
which although providing non-abelian Ramond-Ramond vector mesons, does
not contain a realistic theory. We propose a similar mechanism can be used
to generate the gauge symmetry $SU(3)\otimes SU(2)\otimes U(1)$ from a
symmetric subgroup such as
$SU(2)^2\otimes U(1)^2$.

\vfill\eject
\centerline{\bf Appendix A}
\vskip10pt
\leftline{SIGMA MATRICES}

For the sigma matrices described in (3.1), we have
the following symmetry properties
$$\eqalignno{\sigma^\mu\bar\sigma^\nu = -\sigma^\nu\bar\sigma^\mu
+ 2\eta^{\mu\nu}&\hskip30pt
\bar\sigma^\mu\sigma^\nu = -\bar\sigma^\nu\sigma^\mu
+ 2\eta^{\mu\nu}\cr
\sigma^\mu\bar\sigma^\nu = \eta^{\mu\nu} - {\textstyle{i\over 2}}
\epsilon^{\mu\nu\rho\sigma}\sigma_\rho\bar\sigma_\sigma&\qquad
\bar\sigma^\mu\sigma^\nu = \eta^{\mu\nu} + {\textstyle{i\over 2}}
\epsilon^{\mu\nu\rho\sigma}\bar\sigma_\rho\sigma_\sigma&(A.1)\cr}$$
$$\eqalignno
{\sigma^\mu\bar\sigma^\nu\sigma^\rho + \sigma^\rho\bar\sigma^\nu\sigma^\mu&=
2 (\eta^{\mu\nu}\sigma^\rho -
\eta^{\mu\rho}\sigma^\nu + \eta^{\nu\rho}\sigma^\mu)\cr
\bar\sigma^\mu\sigma^\nu\bar\sigma^\rho +
\bar\sigma^\rho\sigma^\nu\bar\sigma^\mu&=
2 (\eta^{\mu\nu}\bar\sigma^\rho -\eta^{\mu\rho}\bar\sigma^\nu
+\eta^{\nu\rho}\bar\sigma^\mu)
&(A.2)\cr}$$
$$\eqalignno
{\sigma^\mu\bar\sigma^\nu\sigma^\rho - \sigma^\rho\bar\sigma^\nu\sigma^\mu&=
2i\epsilon^{\mu\nu\rho\sigma}\sigma_\sigma \cr
\bar\sigma^\mu\sigma^\nu\bar\sigma^\rho -
\bar\sigma^\rho\sigma^\nu\bar\sigma^\mu&=
-2i\epsilon^{\mu\nu\rho\sigma}\bar\sigma_\sigma&(A.3)\cr}$$
so that
$$\eqalignno
{\sigma^\mu\bar\sigma^\nu\sigma^\rho &=
\eta^{\mu\nu}\sigma^\rho -
\eta^{\mu\rho}\sigma^\nu + \eta^{\nu\rho}\sigma^\mu
+ i \epsilon^{\mu\nu\rho\sigma}\sigma_\sigma\cr
\bar\sigma^\mu\sigma^\nu\bar\sigma^\rho &=
\eta^{\mu\nu}\bar\sigma^\rho -\eta^{\mu\rho}\bar\sigma^\nu
+\eta^{\nu\rho}\bar\sigma^\mu
- i \epsilon^{\mu\nu\rho\sigma}\bar\sigma_\sigma&(A.4)\cr}$$
and the trace properties ${\rm tr}\sigma^\mu\bar\sigma^\nu = 2\eta^{\mu\nu}$,
$$\eqalignno{
{\rm tr}\,\sigma^\rho\bar\sigma^\mu\sigma^\lambda\bar\sigma^\kappa &=
2\eta^{\rho\mu}\eta^{\lambda\kappa} - 2\eta^{\rho\lambda}\eta^{\mu\kappa}
+ 2\eta^{\rho\kappa}\eta^{\mu\lambda} + 2i\epsilon^{\rho\mu\lambda\kappa}\cr
{\rm tr}\,\bar\sigma^\rho\sigma^\mu\bar\sigma^\lambda\sigma^\kappa &=
2\eta^{\rho\mu}\eta^{\lambda\kappa} - 2\eta^{\rho\lambda}\eta^{\mu\kappa}
+ 2\eta^{\rho\kappa}\eta^{\mu\lambda} - 2i\epsilon^{\rho\mu\lambda\kappa}
\,,&(A.5)}$$
and
$$\eqalignno{
{\rm tr}\,\sigma^\rho\bar\sigma^\sigma
\sigma^\mu\bar\sigma^\lambda\sigma^\kappa\bar\sigma^\nu =
2\,[\,&\eta^{\rho\sigma} (\eta^{\mu\lambda}\eta^{\kappa\nu}
-\eta^{\mu\kappa}\eta^{\lambda\nu} + \eta^{\mu\nu}\eta^{\lambda\kappa})\cr
-&\eta^{\rho\mu} (\eta^{\sigma\lambda}\eta^{\kappa\nu}
-\eta^{\sigma\kappa}\eta^{\lambda\nu} +\eta^{\sigma\nu}\eta^{\lambda\kappa})\cr
+&\eta^{\rho\lambda} (\eta^{\sigma\mu}\eta^{\kappa\nu}
-\eta^{\sigma\kappa}\eta^{\mu\nu} + \eta^{\sigma\nu}\eta^{\mu\kappa})\cr
-&\eta^{\rho\kappa} (\eta^{\sigma\mu}\eta^{\lambda\nu}
-\eta^{\sigma\lambda}\eta^{\mu\nu} + \eta^{\sigma\nu}\eta^{\mu\lambda})\cr
+&\eta^{\rho\nu} (\eta^{\sigma\mu}\eta^{\lambda\kappa}
-\eta^{\sigma\lambda}\eta^{\mu\kappa} +\eta^{\sigma\kappa}\eta^{\mu\lambda})\cr
+&i (\eta^{\rho\sigma}\epsilon^{\mu\lambda\kappa\nu}
-\eta^{\sigma\mu}\epsilon^{\lambda\kappa\nu\rho}
-\eta^{\lambda\nu}\epsilon^{\rho\sigma\mu\kappa}\cr
&\hskip3pt -\eta^{\rho\mu}\epsilon^{\sigma\lambda\kappa\nu}
+\eta^{\lambda\kappa}\epsilon^{\rho\sigma\mu\nu}
+\eta^{\kappa\nu}\epsilon^{\rho\sigma\mu\lambda})]\cr
{\rm tr}\,\bar\sigma^\rho\sigma^\sigma
\bar\sigma^\mu\sigma^\lambda\bar\sigma^\kappa\sigma^\nu =
2\,[\,&\eta^{\rho\sigma} (\eta^{\mu\lambda}\eta^{\kappa\nu}
-\eta^{\mu\kappa}\eta^{\lambda\nu} + \eta^{\mu\nu}\eta^{\lambda\kappa})\cr
-&\eta^{\rho\mu} (\eta^{\sigma\lambda}\eta^{\kappa\nu}
-\eta^{\sigma\kappa}\eta^{\lambda\nu} +\eta^{\sigma\nu}\eta^{\lambda\kappa})\cr
+&\eta^{\rho\lambda} (\eta^{\sigma\mu}\eta^{\kappa\nu}
-\eta^{\sigma\kappa}\eta^{\mu\nu} + \eta^{\sigma\nu}\eta^{\mu\kappa})\cr
-&\eta^{\rho\kappa} (\eta^{\sigma\mu}\eta^{\lambda\nu}
-\eta^{\sigma\lambda}\eta^{\mu\nu} + \eta^{\sigma\nu}\eta^{\mu\lambda})\cr
+&\eta^{\rho\nu} (\eta^{\sigma\mu}\eta^{\lambda\kappa}
-\eta^{\sigma\lambda}\eta^{\mu\kappa} +\eta^{\sigma\kappa}\eta^{\mu\lambda})\cr
-&i (\eta^{\rho\sigma}\epsilon^{\mu\lambda\kappa\nu}
-\eta^{\sigma\mu}\epsilon^{\lambda\kappa\nu\rho}
-\eta^{\lambda\nu}\epsilon^{\rho\sigma\mu\kappa}\cr
&\hskip3pt -\eta^{\rho\mu}\epsilon^{\sigma\lambda\kappa\nu}
+\eta^{\lambda\kappa}\epsilon^{\rho\sigma\mu\nu}
+\eta^{\kappa\nu}\epsilon^{\rho\sigma\mu\lambda})]\,.
&(A.5)\cr}$$
\vskip15pt
\centerline{\bf Appendix B}
\vskip10pt
\leftline{GAMMA MATRICES}

Higher-dimensional gamma matrices occur naturally in critical superstring
theory. In (3.1) and (4.3) we have given a Weyl representation for the
four and six-dimensional cases, respectively. In this appendix,
we consider in ten dimensions the $\Gamma$ matrices satisfying
$\{\Gamma^M,\Gamma^N\} = 2\eta^{MN}$;
$\eta^{MN} = {\rm diag} \{-1, 1, \ldots, 1\}$.
In a Weyl representation, they have only off-diagonal components and
can be written as
$$\Gamma^\mu = \left(\matrix{0&I_4\cr
I_4&0\cr}\right)\otimes\gamma^\mu\,;\quad
\Gamma^{a+3}= \left(\matrix{0&\alpha^a\cr
-\alpha^a&0\cr}\right)\otimes I_4\,;\quad
\Gamma^{a+6}= \left(\matrix{0&\beta^a\cr
\beta^a&0}\right)\otimes\bar\gamma^5\,.\eqno(B.1)$$
Here $0 \le M,N \le 9$; $0\le\mu\le 3$ and $1\le a \le 3$; and
$\{\gamma^\mu\,,\,\gamma^\nu\} = 2 \eta^{\mu\nu}$.
The six matrices $\alpha^a$, $\beta^a$ are antisymmetric and real and satisfy
the following algebra:
$$\{\alpha^a,\alpha^b\} = \{\beta^a,\beta^b\} = -2\delta^{ab}$$
$$[\alpha^a,\beta^b ] = 0\,,\quad
[\alpha^a,\alpha^b ] = -2\epsilon_{abc}\alpha^c\,,\quad
[\beta^a,\beta^b ] = 2\epsilon_{abc}\beta^c\,.\eqno(B.2)$$
An explicit representation of these matrices is given in terms of the Pauli
matrices:
$$\beta^1 =\left(\matrix{0&i\sigma^2\cr
i\sigma^2&0\cr}\right)\,,\quad
\beta^2 =\left(\matrix{0&1\cr
-1&0\cr}\right)\,,\quad
\beta^3 =\left(\matrix{-i\sigma^2&0\cr
0&i\sigma^2\cr}\right)\,,$$
$$\alpha^1 =\left(\matrix{0&\sigma^1\cr
-\sigma^1&0\cr}\right)\,,\quad
\alpha^2 =\left(\matrix{0&-\sigma^3\cr
\sigma^3&0\cr}\right)\,,\quad
\alpha^3 =\left(\matrix{i\sigma^2&0\cr
0&i\sigma^2\cr}\right)\,.\eqno(B.3)$$
The hermitian chirality operator is
$\Gamma^{11} \equiv \Gamma^0\ldots\Gamma^9 =
\left(\matrix{I_4&0\cr
0&-I_4\cr}\right)\otimes I_4\,, \quad(\Gamma^{11})^2 = 1$ and
$(\bar\gamma^5)^2= - 1$, so $\bar\gamma^5$ is antihermitian and
is given by $\gamma^5 = i\gamma^0\gamma^1\gamma^2
\gamma^3 = i\bar\gamma^5$.
The matrix direct product is
$A\otimes B \equiv \left(\matrix{a_{11}B&a_{12}B&...\cr
a_{21}B&a_{22}B&...\cr
...&\cr}\right)$; it follows that
$(A\otimes B) (C\otimes D) = AC\otimes BD$.
We denote the index structure of the $\Gamma$ matrices as
$\Gamma^{\mu{\cal A}}_{\hskip 9pt{\cal B}}$ and tensors which raise and
lower spinor indices are the antisymmetric tensors
$C^{-1}_{\cal AB}$, $C^{\cal AB}$, the charge conjugation matrices for
$SO(9,1)$. As before, $\chi_{\cal A} = C^{-1}_{\cal AB}\chi^{\cal B}$,
$\chi^{\cal A} = C^{\cal AB}\chi_{\cal B}$,
$C^{-1}_{\cal AB}C^{\cal BC} =\delta_{\cal A}^{\cal C}$,
$C^{\cal AB} = - C^{\cal BA}$,
$C^{-1}_{\cal AB} = - C^{-1}_{\cal BA}$.
Then $A_{\cal A} B^{\cal A} = -A^{\cal A} B_{\cal A}$.
It follows from the definition of the charge conjugation matrix
$C^{-1}_{\cal AB} (\Gamma^\mu)^{\cal B}\hskip1pt_{\cal E}C^{\cal ED} =
-(\Gamma^{\mu T})_{\cal A}\hskip1pt^{\cal D}$ that
$(\Gamma^\mu)^{\cal D}\hskip1pt_{\cal A}=
(\Gamma^\mu)_{\cal A}\hskip1pt^{\cal D}\,;$
and that $\Gamma^{\mu{\cal AB}}$ and $\Gamma^\mu_{\cal AB}$
are symmetric in the spinor indices.
The exact form of $C$ depends on the representation of the $\Gamma$ matrices
used. From (B.1), we have
$$C = \left(\matrix{0&I_4\cr
I_4&0\cr}\right)\otimes C_4\,\quad ;\qquad
C^{-1} = \left(\matrix{0&I_4\cr
I_4&0\cr}\right)\otimes C_4^{-1}\,\eqno(B.4)$$
where $C_4$ is the charge conjugation matrix for the four-dimensional
$\gamma$ matrices: $C_4^{-1}(\gamma^\mu)C_4 =
-(\gamma^{\mu T})$ and thus $\gamma^{5*} = C^{-1} \gamma^5 C\,$.
We will denote $1\le{\cal A}\le 32$ as $1\le A,\dot A\le 16$.
\vskip15pt
\centerline{\bf References}
\vskip 5pt
\item{ 1.} D. Friedan, E. Martinec, and S. Shenker, Nucl. Phys. {\bf B271}
(1986) 93.
\item{ 2.} D. Lust and S. Theisen, {\it Lectures on String Theory}.
New York: Springer-Verlag, 1989.
\item{ 3.} D. Friedan, {\it Workshop on Unified String Theories},
Green, M. and Gross, D., eds., Singapore: World Scientific,
1986, pp. 162-213.
\item{ 4.} B. L. van der Waerden, {\it Group Theory and Quantum Mechanics}.
New York: Springer Verlag, (1974).
\item{ 5.} J. Wess and J. Bagger, {\it Supersymmetry and Supergravity}.
Princeton: Princeton University Press (1992).
\item{ 6.} L. Dolan and S. Horvath, Nucl. Phys. {\bf B416} (1994) 87.
\item{ 7.} J. Scherk and J. Schwarz, Nucl. Phys. {\bf B81} (1974) 118.
\item{ 8.} F. Gliozzi, J. Scherk and D. Olive,
Nucl. Phys. {\bf B122} (1977) 253.
\item{ 9.} L. Dolan, P. Goddard and P. Montague, hep-th/9410029 preprint.
\item{10.} A. Belavin, A. Polyakov, and A.B. Zamolodchikov,
Nucl. Phys. {\bf B241} (1984) 333.
\item{11.} V. Knizhnik and A.B. Zamolodchikov, Nucl. Phys. {\bf B247}
(1984) 83.
\item{12.} D. Lust, S. Theisen and G. Zoupanos, Nucl. Phys. {\bf B296}
(1988) 800.
\item{13.} R. Bluhm, L. Dolan and P. Goddard, Nucl. Phys.
{\bf B338} (1990) 529.
\item{14.} J. Harvey, A. Strominger, and C. Callan, Nucl. Phys.
{\bf B367} (1991) 60; M. Duff and  J.X. Lu,  Nucl. Phys. {\bf B416} (1995) 301.
\item{15.} R. Bluhm, L. Dolan and P. Goddard, Nucl. Phys.
{\bf B289} 364 (1987).
\item{16.} L. Dixon, V. Kaplunovsky and C. Vafa, Nucl. Phys.
{\bf B294} (1987) 43.
\item{17.} N. Berkovits and C. Vafa,
Mod. Phys. Lett. {\bf A9} (1994) 653: hep-th/9310170.
\item{18.} T. Banks, J. Dixon, D. Friedan, and E. Martinec,
Nucl. Phys. {\bf B299} (1988) 613.
\item{19.} T. Banks, J. Dixon, Nucl. Phys. {\bf B307} (1988) 93.
\item{20.} W. Lerche, D. Lust and A.N. Schellekens, Nucl. Phys. {\bf B287}
(1987) 477.
\item{21.} C. Thorn, Phys. Rev. {\bf D4} (1971) 1112;
J. Schwarz, Phys. Lett. {\bf B37} (1971) 315;
E. Corrigan and D. Olive, Nuovo Cim. {\bf 11A} (1971) 749;
E. Corrigan and P. Goddard, Nuovo Cim. {\bf 18A} (1973) 339.
\item{22.} M.B. Green, J. Schwarz and E. Witten, {\it Superstring
theory} (Cambridge University Press, 1987).
\item{23.} L. Dolan and S. Horvath, in preparation.
\item{24.} C. Hull and P. Townsend, hep-th/9410167.
\vfill\eject

\bye